\documentstyle[12pt]{article}
\begin{document}
\begin{flushright}
hep-th/0609201\\ CAS-BHU/Preprint
\end{flushright}
\vskip 2cm
\begin{center}
{\bf \large {ONE-FORM ABELIAN GAUGE THEORY AS THE HODGE THEORY}}

\vskip 2cm

{\bf R.P.Malik}\\
{\it Centre of Advanced Studies, Physics
Department,}\\ {\it Banaras Hindu University, Varanasi-221 005,
India}
\\ {\bf E-mail address: malik@bhu.ac.in}

\vskip 1.5cm

\end{center}

\noindent {\bf Abstract}: We demonstrate that the two (1 +
1)-dimensional (2D) free 1-form Abelian gauge theory provides an
interesting field theoretical model for the Hodge theory. The
physical symmetries of the theory correspond to all the basic
mathematical ingredients that are required in the definition of
the de Rham cohomological operators of differential geometry. The
conserved charges, corresponding to the above continuous symmetry
transformations, constitute an algebra that is reminiscent of the
algebra obeyed by the de Rham cohomological operators. The
topological features of the above theory are discussed in terms of
the BRST and co-BRST operators. The super de Rham cohomological
operators are exploited in the derivation of the nilpotent
(anti-)BRST, (anti-)co-BRST symmetry transformations and the
equations of motion for all the fields of the theory, within the
framework of the superfield formulation. The derivation of the
equations of motion, by exploiting the super Laplacian operator,
is a completely new result in the framework of the superfield
approach to BRST formalism. In an Appendix, the interacting 2D
Abelian gauge theory (where there is a coupling between the U(1)
gauge field and the Dirac fields) is also shown to provide a
tractable field theoretical model for the Hodge theory.\\

\baselineskip=16pt

\vskip .5cm



\newpage

\noindent
{\bf 1 Introduction}\\

\noindent There are a host of areas of research in theoretical
high energy physics which have provided a meeting-ground for the
researchers in the realm of mathematics and investigators in the
domain of theoretical high energy physics. Such areas of
investigations have enriched the understanding and insights of
both the above type of researchers in an illuminating and fruitful
fashion. The subject of Becchi-Rouet-Stora-Tyutin (BRST) formalism
[1-4] is one such area of research which has found applications in
the modern developments in (super)string theories, D-branes,
M-theory, etc., that are supposed to be the frontier areas of
research in modern-day theoretical high energy physics (see, e.g.
[5,6]). In particular, in the context of string field theories,
the BRST formalism plays a very decisive role.

One of the most important pillars of strength for the edifice of
the ideas behind the BRST formalism is its very fruitful
application in the realm of 1-form (non-)Abelian gauge theories in
the physical four (3 + 1)-dimensions of spacetime. The latter
theories provide the physical basis for the existence of three out
of four fundamental interactions of nature. In fact, the above
gauge theories are described by the singular Lagrangian densities
and they are endowed with the first-class constraints in the
language of Dirac's prescription for the classification scheme of
the constraints [7,8]. For such dynamical systems with the
first-class constraints, the BRST formalism provides (i) the
covariant canonical quantization where the ``classical'' local
gauge symmetry transformations of the original theory are traded
with the ``quantum'' gauge (i.e. BRST) symmetry transformations,
(ii) the physical mechanism for the proof of unitarity at any
arbitrary given order of perturbative computations for a physical
process allowed by the theory [9,10], and (iii) the method to
choose the physical states from the total quantum Hilbert space of
states which are found to be consistent with the Dirac's
prescription for the quantization scheme of the systems with
constraints.

The range and reach of the BRST formalism has been extended to
incorporate in its ever widening folds the second-class
constraints, too [11]. Its very deep connections with the key
notions of the differential geometry and cohomology [12-16], its
beautiful application in the context of topological field theories
[17-19], its very intimate relations with the key ideas behind the
supersymmetry, its cute geometrical origin and interpretation in
the framework of the superfield formulation [20-29], its
successful applications to the reparametrization invariant
theories of free as well as interacting relativistic particle,
supergravity, etc., have elevated the key concepts behind the BRST
formalism to a very high degree of mathematical sophistication and
very useful physical applications.

In our present investigation, we shall concentrate on the
application of the BRST formalism to the 2D free 1-form Abelian
gauge theory (which is endowed with the first-class constraints).
In the BRST approach to any arbitrary $p$-form ($p = 1, 2, 3....$)
gauge theories, the operator form of the first class constraints
appear in the physicality condition (i.e. $Q_b |phys> = 0$) when
one demands that the physical states (i.e. $|phys>$) of the
quantum gauge theories are annihilated by the nilpotent ($Q_b^2 =
0$) and conserved ($\dot Q_b = 0$) BRST charge operator $Q_b$. The
nilpotency of the BRST charge and the physicality condition are
the two key ingredients that allow the BRST formalism to have
close connections with the key ideas of the mathematical aspects
of differential geometry and cohomology. In fact, two physical
(i.e. $Q_b |phys>^\prime = Q_b |phys> = 0$) states $|phys>$ and
$|phys>^\prime = |phys> + Q_b\; |\xi>$ (for $|\xi>$ being a
non-trivial state) are said to belong to the same cohomology class
with respect to the BRST charge $Q_b$ if they differ by a BRST
exact (i.e. $Q_b |\xi>$) state.

In the language of the differential geometry and differential
forms, two closed (i.e. $d f^\prime_n = 0, d f_n = 0$) forms $f_n$
and $f^\prime_n = f_n + d g_{n - 1}$ of degree $n$ (with $n = 1,
2, 3.....)$ are said to belong to the same cohomology class with
respect to the exterior derivative $d = dx^\mu
\partial_\mu$ (with $d^2 = 0$) because they differ from each-other
by an exact form $d g_{n - 1}$. Thus, we note that the nilpotent
BRST charge $Q_b$, that generates a set of nilpotent BRST
transformations for the appropriate and relevant fields of the
gauge theories, provides a physical analogue to the mathematically
abstract cohomological operator $d = dx^\mu\partial_\mu$. There
are two other cohomological operators $\delta = \pm * d *$ and $
\Delta = d \delta + \delta d \equiv \{ d, \delta \}$  (where $*$
is the Hodge duality operation) that constitute the full set ($d,
\delta, \Delta)$ of the de Rham cohomological operators obeying
the algebra $d^2 = \delta^2 = 0, \Delta = (d + \delta)^2 \equiv
\{d, \delta\}, [\Delta, d] = [\Delta, \delta ] = 0$ (see, e.g.,
[12-14] for details). The latter two cohomological operators
$\delta$ and $ \Delta$ are known as the co-exterior derivative and
the Laplacian operator, respectively, in the domain of
differential geometry.

In terms of the above cohomological operators any n-form $f_n$ can
be uniquely written as the sum of a harmonic form $h_n$ (with
$\Delta h_n = 0, d h_n = 0, \delta h_n = 0$) an exact form $d e_{n
- 1}$ and a co-exact form $\delta c_{n + 1}$ on a compact manifold
without a boundary. Mathematically, this statement can be
succinctly expressed as (see, e.g. [12-16] for details)$$
\begin{array}{lcl}
f_n = h_n + d \; e_{n - 1} + \delta\; c_{n + 1}.
\end{array}\eqno(1.1)
$$ The above equation is the statement of the celebrated Hodge
decomposition theorem on the compact manifold without a boundary.
It has been a long-standing problem, in the framework of the BRST
formalism, to obtain the analogue of the cohomological operators
$\delta$ and $\Delta$ in the language of the well-defined symmetry
properties of the Lagrangian density of any arbitrary $p$-form ($p
= 1, 2 ....$) gauge theory in any arbitrary dimension of
spacetime. Some attempts [30-34], in this direction, have been
made for the physical four (3 + 1)-dimensional (4D) (non-)Abelian
1-form gauge theories but the symmetry transformations turn out to
be non-local and non-covariant. In the covariant formulation of
the above symmetry transformations [35], the nilpotency of the
transformations is restored {\it only} for a specific value of a
parameter (that is introduced by hand in the covariant
formulation).

In our earlier set of papers [36-43], we have been able to
demonstrate that (i) the 2D {\it free} as well as {\it
interacting} Abelian 1-form gauge theory [36,38-40], (ii) the
self-interacting 2D non-Abelian 1-form gauge theory without any
interaction with matter fields [37], and (iii) the free 4D Abelian
2-form gauge theory [42-44], provide the tractable field
theoretical models for the Hodge theory because all the de Rham
cohomological operators $(d, \delta, \Delta)$ are shown to
correspond to local, covariant and continuous symmetry
transformations. The discrete set of symmetry transformations for
the Lagrangian densities of the above theories are shown to
correspond to the Hodge duality $*$ operation of the differential
geometry. In fact, the interplay between the continuous and
discrete symmetry transformations provides the analogue of the
relationship $\delta = \pm * d *$ that exists in the differential
geometry. The point to be emphasized here is the fact that all the
above symmetry transformations turn out to be well-defined. In
other words, all the symmetry transformations corresponding to the
cohomological operators are {\it not} found to be non-local or
non-covariant for the above theories. The topological features of
the above 2D and 4D theories are also discussed, in great detail,
by exploiting the nilpotent (anti-)BRST and (anti-)co-BRST
symmetry transformations (and their corresponding nilpotent
generators) {\it together} [41-43].

It is a well-known fact that, for a given single set of local
gauge symmetry transformation for a 1-form (non-)Abelian gauge
theory, there exist {\it two} sets of nilpotent symmetry
transformations (and their corresponding nilpotent generators).
These nilpotent symmetry transformations are known as the
(anti-)BRST symmetry transformations $s_{(a)b}$ which are
generated by the conserved and nilpotent (anti-)BRST charges
$Q_{(a)b}$. The theoretical reason behind the existence of these
couple of nilpotent symmetry transformations, corresponding to a
single local gauge symmetry transformation, comes from the super
exterior derivative $\tilde d = dx^\mu
\partial_\mu + d \theta \partial_\theta + d \bar\theta
\partial_{\bar\theta}$ (with $\tilde d^2 = 0$) when it is
exploited in the celebrated horizontality condition [20-29] on a
(D, 2)-dimensional supermanifold on which a D-dimensional 1-form
(non-)Abelian gauge theory is considered (see, e.g., Subsec. 4.1
below, for details). This technique, popularly known as the
superfield approach to BRST formalism, sheds light on the
geometrical origin and interpretation for the (anti-)BRST symmetry
transformations ($s_{(a)b}$) and corresponding nilpotent
generators ($Q_{(a)b}$). One of the outstanding problems in the
superfield approach to BRST formalism has been to tap the
potential and power of the super co-exterior derivative $\tilde
\delta = \pm \star \tilde d \star$ and the super Laplacian
operator $\tilde \Delta = \tilde d \tilde \delta + \tilde \delta
\tilde d$ in the derivation of some physically relevant aspects of
a $p$-form (non-)Abelian gauge theory in D-dimensions of
spacetime. In the above, it will be noted that the Hodge duality
$\star$ operation is defined on the (D, 2)-dimensional
supermanifold.

A {\it perfect} field theoretical model for the Hodge theory is
the one where (i) the analogues of the {\it ordinary} de Rham
cohomological operators exist in the language of the well-defined
symmetry transformations (and their corresponding generators), and
(ii) all the {\it super} de Rham cohomological operators play very
significant and decisive roles in the determination of some of the
key features of the field theoretical model. The central purpose
of the present investigation is to demonstrate that the 2D free
1-form Abelian gauge theory is a tractable field theoretical model
for the Hodge theory because {\it both} the above {\it key}
requirements are fulfilled in a grand and illuminating manner.
Moreover, we also mention the physical consequences of our
theoretical study in a concise manner. Thus, in our present paper,
the mathematical and physical aspects of our 2D 1-form gauge model
have been brought {\it together} in a cute and complete fashion.
First of all, in Sec. 2, we demonstrate that, corresponding to
each cohomological operators of the differential geometry, there
exists a well-defined symmetry transformation for the Lagrangian
density of the 2D free 1-form Abelian gauge theory. One of the
physical consequences of the above symmetry transformations is the
fact that, the gauge theory under consideration, is found to be a
{\it new} type of topological field theory (cf. Sec. 3 below).
Parallel to Sec. 2, we demonstrate (in Sec. 4) that the {\it
super} de Rham cohomological operators ($\tilde d, \tilde \delta =
- \star \tilde d \star, \tilde \Delta = \tilde d \tilde \delta +
\tilde \delta \tilde d)$ play central roles in the appropriate
(gauge-invariant) restrictions on the superfield of four (2,
2)-dimensional supermanifold which generate (i) the well-defined
nilpotent (anti-)BRST symmetry transformations, (ii) the
well-defined nilpotent (anti-)co-BRST symmetry transformations,
and (iii) the equations of motion for all the fields of the
theory. In the application of the super cohomological operators
$\tilde \delta$ and $\tilde \Delta$, we require a proper
definition of the Hodge duality $\star$ operation on the four (2,
2)-dimensional supermanifold.


In the language of the symmetry properties of the Lagrangian
density (cf. (2.2) below), we have been able to provide the
analogue of the Hodge duality $*$ operation that exists between
the exterior derivative $d$ and the co-exterior derivative
$\delta$ in the well-known relationship $\delta = - * d *$ of the
differential geometry. In fact, the discrete symmetry
transformations (cf. (2.10) and (2.11) below) for the Lagrangian
density of the theory plays the role of the Hodge duality $*$
operation in the relationships (cf. (2.13) and (2.14) below) that
exists between the (anti-)co-BRST and (anti-)BRST symmetry
transformations. However, we also know that there exists a
well-defined meaning of the Hodge duality $*$ operation on the
differential forms (through their proper definition of the inner
products) on the 2D spacetime manifold (see, e.g. [12-16]). To
obtain the analogy of this Hodge duality $*$ operation (defined on
an ordinary spacetime manifold), the key point is to know the
proper definition of the corresponding Hodge duality $\star$
operation on the super differential forms defined on the four (2,
2)-dimensional supermanifold. We have achieved precisely this goal
in our earlier work [45]. The materials of our Subsecs. 4.2 and
4.3, where we have exploited the super co-exterior derivative
$\tilde \delta = - \star \tilde d \star$ and super Laplacian
operator $\tilde \Delta$ in a specific set of restrictions on the
superfields of the above supermanifold, totally depend on the
definition of the Hodge duality $\star$ operation on the super
forms [45]. The results of the Hodge duality  $\star$ operation
are found to be correct.

Our present investigation is essential and interesting on the
following grounds. First and foremost, our present field
theoretical model is one of the simplest examples where the
sanctity of our definition of the Hodge duality $\star$ operation
on the four (2, 2)-dimensional supermanifold [45] can be tested,
particularly, in the application of the super co-exterior
derivative $\tilde \delta = - \star \tilde d \star$ and the super
Laplacian operator $\tilde \Delta = \tilde d \tilde \delta +
\tilde \delta \tilde d$ in some suitable restrictions on the
superfields of the above supermanifold . Second, the present model
provides the physical meaning of (and theoretical importance to)
the {\it ordinary} and {\it super} de Rham cohomological operators
{\it together}. The physical implication of the former lies in the
proof that the present model is a {\it new} kind of TFT. The
theoretical importance of the latter cohomological operators is in
the derivation of the nilpotent symmetry transformations and
equations of motion for the theory. Third, due to the aesthetic
reasons, it is nice to note that, for the model under
consideration, the continuous and discrete symmetries,
mathematical power of the cohomological operators and their
physical consequences, etc., are found to blend together in a
beautiful manner. Finally, the present study is a step in the
direction to prove that the free 2-form Abelian gauge theory might
provide a field theoretical model for the Hodge theory in the
physical four dimensions of spacetime where the ordinary as well
as the super de Rham cohomological operators would play
significant and decisive roles. The physical implication of the
former operators has already been shown in the proof that the 4D
free 2-form Abelian gauge theory is endowed with some special
features and is a model for the quasi-topological field theory
[43]. The impact and importance of the latter (i.e. super de Rham
cohomological operators) are yet to be seen in the context of
theoretical discussions of the above 4D free 2-form Abelian gauge
theory.

The contents of our present paper are organized as follows:

In Sec. 2, we discuss the bare essentials of (i) the nilpotent
(anti-)BRST symmetry, (ii) the nilpotent (anti-)co-BRST symmetry,
and (iii) a non-nilpotent bosonic symmetry transformations for the
2D (anti-)BRST invariant Lagrangian density of a free 1-form
Abelian gauge theory. The subtle discrete symmetry transformations
for the above Lagrangian density are discussed separately and
independently. We pinpoint the deep connections that exist between
these (continuous and discrete)  symmetry transformations and the
cohomological operators of the differential geometry. This
exercise provides, at a very elementary level, the proof that the
above Abelian gauge theory, described in terms of the BRST
invariant Lagrangian density, is a tractable field theoretical
model for the Hodge theory.

Section 3 is devoted to demonstrate that the above 1-form gauge
theory is a new type of topological field theory which captures  a
part of the key features associated with the Witten-type
topological field theory as well as a part of the salient points
connected with the Schwarz-type topological field theory. The
existence of the nilpotent (anti-)BRST as well as (anti-)co-BRST
symmetry transformations (and their corresponding nilpotent
generators) play a pivotal role in this proof. We do not discuss
here the topological invariants and their recursion relations
which can be found in our earlier works [41,36].

In Sec. 4, we exploit the mathematical power of the super de Rham
cohomological operators, in the imposition of some specific
(gauge-invariant) restrictions on the superfields of the four (2,
2)-dimensional supermanifold, to derive the nilpotent (anti-)BRST
and (anti-)co-BRST symmetry transformations of the theory. We show
that super Laplacian operator plays a key role in the derivation
of all the equations of motion for all the fields of the theory.
This latter result is a completely new result which bolsters up
the correctness of our definition of the Hodge duality $\star$
operation on the above supermanifold (see. e.g. [45] for details).
The topological features of the above theory are also captured in
the language of the superfield approach to BRST formalism. We have
focused, for the proof of the topological features, on the form of
the Lagrangian density and the symmetric energy-momentum tensor
expressed in terms of the superfields defined on the four (2,
2)-dimensional supermanifold.

Finally, we summarize our key results, make some concluding
remarks and point out some promising future directions for further
investigations in Sec. 5.

In the Appendix, we show that the 2D interacting 1-form Abelian
U(1) gauge theory with Dirac fields is a cute field theoretical
model of the celebrated Hodge theory.\\

\noindent {\bf 2 Cohomological Operators and Symmetries:
Lagrangian Formulation}\\

\noindent To establish the connection between the key concepts
behind the de Rham cohomological operators of the differential
geometry and the symmetry properties of the (anti-)BRST invariant
Lagrangian density of a given two (1 + 1)-dimensional\footnote{We
follow here the conventions and notations such that the flat 2D
Minkowskian metric $\eta_{\mu\nu} = $ diag $(+1, -1)$ and the
antisymmetric Levi-Civita tensor $\varepsilon_{01} = + 1 = -
\varepsilon^{01}$ with $\varepsilon^{\mu\nu} \varepsilon_{\mu\nu}
= - 2!, \varepsilon^{\mu\lambda} \varepsilon_{\mu\nu} = -
\delta^{\lambda}_\nu$ etc. Here the Greek indices $\mu, \nu,
\lambda......= 0, 1$ stand for the time and space directions on
the 2D Minkowskian spacetime manifold, respectively. All the local
fields of the 2D free 1-form Abelian gauge theory are defined on
this spacetime manifold because they are functions of the 2D
spacetime variable $x^\mu$.} (2D) free 1-form Abelian gauge
theory, we begin with the following Lagrangian density in the
Feynman gauge [46,47] $$
\begin{array}{lcl}
{\cal L}_b = - {\displaystyle \frac{1}{4} F^{\mu\nu} F_{\mu\nu} -
\frac{1}{2} (\partial \cdot A)^2 - i \;\partial_\mu \bar C
\partial^\mu C} \;\equiv\;
 {\displaystyle \frac{1}{2} \; E^2 - \frac{1}{2} (\partial \cdot
 A)^2 - i\; \partial_\mu \bar C \partial^\mu C},
\end{array} \eqno(2.1)
$$ where $F_{\mu\nu} = \partial_\mu A_\nu - \partial_\nu A_\mu$ is
the field strength tensor for the 2D Abelian 1-form ($A^{(1)} =
dx^\mu A_\mu$) gauge field $A_\mu$. It has only one-component
$F_{01} = \partial_0 A_1 - \partial_1 A_0 = E$ which turns out to
be the electric field $E$ of the theory. There exists no magnetic
field in the 2D Abelian gauge theory. The cohomological origin for
the existence of the field strength tensor lies in the exterior
derivative $d = dx^\mu \partial_\mu$ (with $d^2 = 0$) because the
2-form $F^{(2)} = d A^{(1)} = \frac{1}{2} (dx^\mu \wedge dx^\nu)
F_{\mu\nu}$, constructed with the help of $d$ and $A^{(1)}$,
defines it. On the other hand, the gauge-fixing term $(\partial
\cdot A)$ owes its cohomological origin to the co-exterior
derivative $\delta = - * d *$ (with $\delta^2 = 0$) because
$\delta A^{(1)} = - * d * A^{(1)} = (\partial \cdot A)$ where $*$
is the Hodge duality operation defined on the 2D Minkowskian
spacetime manifold. The (anti-)ghost fields $(\bar C)C$ are
required in the theory for the proof of unitarity for a given
physical process at any arbitrary given order of the perturbative
computation\footnote{The importance of the fermionic
(anticommuting) (anti-)ghost fields, in the proof of unitarity for
a physical process, comes out in its full blaze of glory in the
context of the non-Abelian gauge theory where for each gluon loop
diagram (that exists for a given physical process), one requires a
loop Feynman diagram constructed with the help of the fermionic
(anti-)ghost fields {\it alone} (see, e.g. [9,10] for details).}.
For the 1-form Abelian gauge theory, the ``fermionic''
(anti-)ghost fields are (i) the spin-zero Lorentz scalar fields,
and (ii) they possess anticommuting (i.e. $C^2 = \bar C^2 = 0, C
\bar C + \bar C C = 0$) property.

The square terms (i.e. $\frac{1}{2} E^2, - \frac{1}{2} (\partial
\cdot A)^2 $) corresponding to the kinetic energy term and the
gauge-fixing term can be linearized by introducing the auxiliary
fields $B$ and ${\cal B}$ thereby changing the above Lagrangian
density (2.1) in the following equivalent form $$
\begin{array}{lcl}
{\cal L}_B =  {\displaystyle {\cal B}\; E - \frac{1}{2}\; {\cal
B}^2 + B \;(\partial \cdot A) + \frac{1}{2}\; B^2 - i\;
\partial_\mu \bar C
\partial^\mu C},
\end{array} \eqno(2.2)
$$ where the auxiliary field $B$ is popularly known as the
Nakanishi-Lautrup auxiliary field. The above Lagrangian density is
endowed with the following off-shell nilpotent ($s_{(a)b}^2 = 0$)
and anticommuting ($s_b s_{ab} + s_{ab} s_b = 0$) (anti-)BRST
symmetry transformations $s_{(a)b}$\footnote{We follow here the
notations and conventions adopted in [47]. In fact, in its
totality, the BRST transformation $\delta_B$ is a product
($\delta_B = \eta s_b$) of an anticommuting (i.e. $\eta C + C \eta
= 0, \eta \bar C + \bar C \eta = 0$) spacetime independent
parameter $\eta$ and the nilpotent ($s_b^2 = 0$) operator $s_b$.}
[46,47] $$
\begin{array}{lcl}
&&s_b A_\mu = \partial_\mu C, \qquad s_b C = 0, \qquad s_b \bar C
= i B, \qquad s_b B = 0, \nonumber\\ && s_b E = 0, \qquad s_b
{\cal B} = 0, \qquad s_b (\partial \cdot A) = \Box C, \qquad s_b
F_{\mu\nu} = 0, \nonumber\\ &&s_{ab} A_\mu = \partial_\mu \bar C,
\qquad s_{ab} \bar C = 0, \qquad s_{ab} C = - i B, \qquad s_{ab} B
= 0, \nonumber\\ && s_{ab} E = 0, \qquad s_{ab} {\cal B} = 0,
\qquad s_{ab} (\partial \cdot A) = \Box \bar C, \qquad s_{ab}
F_{\mu\nu} = 0,
\end{array} \eqno(2.3)
$$ because the above Lagrangian density transforms as: $s_b {\cal
L}_B = \partial_\mu [B \partial^\mu C]$ and $s_{ab} {\cal L}_B =
\partial_\mu [B \partial^\mu \bar C]$, respectively, under the
nilpotent transformations (2.3). It will be noted that the gauge
invariant physical field $E$ remains invariant under the nilpotent
(anti-)BRST transformations listed in (2.3). We know, however,
that the cohomological origin for the above electric field $E$ is
encoded in the exterior derivative $d = dx^\mu \partial_\mu$ which
generates the 2-form $F^{(2)} = d A^{(1)}$. The latter, in turn,
produces the field strength tensor $F_{\mu\nu} = \partial_\mu
A_\nu -
\partial_\nu A_\mu$. Thus, we conclude that the mathematical origin
of the nilpotent (anti-)BRST symmetry transformations (e.g. for
our present 2D 1-form free Abelian gauge theory) lies in the
exterior derivative $d = dx^\mu \partial_\mu$ of the differential
geometry. This observation will be exploited in Subsec. 4.1 where
the super exterior derivative, exploited in the so-called
horizontality condition [20-29], will generate the nilpotent
(anti-)BRST symmetry transformations {\it together} for all the
fields of the above 1-form Abelian gauge theory in the framework
of the geometrical superfield approach to BRST formalism.

The Lagrangian density (2.2) is found to be endowed with another
off-shell nilpotent (i.e. $s_{(a)d}^2 = 0$) symmetry
transformations. The latter transformations are christened as the
dual(co-) and anti-dual(co-)BRST symmetry transformations
$s_{(a)d}$. In fact, it can be checked that, under the following
(anti-)co-BRST symmetry transformations [36,38,41] $$
\begin{array}{lcl}
 &&s_d A_\mu = - \varepsilon_{\mu\nu}
\partial^\nu \bar C, \qquad s_d \bar C = 0, \qquad s_d  C = - i {\cal B}, \qquad
s_d {\cal B} = 0, \nonumber\\ && s_d E = \Box \bar C, \quad s_d B
= 0, \quad s_d (\partial \cdot A) = 0, \quad s_d F_{\mu\nu} =
[\varepsilon_{\mu\rho} \partial_\nu - \varepsilon_{\nu\rho}
\partial_\mu]\; \partial^\rho \bar C, \nonumber\\
&&s_{ad} A_\mu = - \varepsilon_{\mu\nu}
\partial^\nu  C, \qquad s_{ad}  C = 0, \qquad s_{ad} \bar C =
 i {\cal B}, \qquad s_{ad} {\cal B} = 0, \nonumber\\ && s_{ad} E
= \Box C, \quad s_{ad}  B = 0, \quad s_{ad} (\partial \cdot A) =
0, \quad s_{ad} F_{\mu\nu} = [\varepsilon_{\mu\rho} \partial_\nu -
\varepsilon_{\nu\rho}
\partial_\mu]\; \partial^\rho  C,
\end{array}\eqno(2.4)
$$ (i) the Lagrangian density (2.2) for the 2D free 1-form Abelian
theory changes to a total derivative (i.e. $s_d {\cal L}_B =
\partial_\mu [{\cal B} \partial^\mu \bar C],\; s_{ad} {\cal L}_B =
\partial_\mu [ {\cal B} \partial^\mu C]$),
\\ (ii) the anticommuting nature of the nilpotent (anti-)co-BRST
symmetry transformations becomes transparent because the operator
equation  $[s_d s_{ad} + s_{ad} s_d] = 0$ turns out to be true for
any arbitrary field $\Omega$ (i.e. $[s_d s_{ad} + s_{ad} s_d]\;
\Omega = 0$) of the Lagrangian density (2.2),\\ (iii) the
gauge-fixing term $(\partial \cdot A)$, owing its origin to the
nilpotent co-exterior derivative $\delta = - * d *$, remains
invariant under the above (anti-)co-BRST symmetry
transformations,\\ (iv) the gauge-fixing term $(\partial \cdot A)$
is an on-shell (i.e. $\Box C = 0, \Box \bar C = 0$) invariant
quantity under the nilpotent (anti-)BRST transformations (2.3),
and\\ (v) the cohomological origin for the existence of the
(anti-)co-BRST symmetry transformations for the above gauge theory
lies in the dual(co) exterior derivative $\delta$. This
observation will play an important role in the derivation of the
(anti-)co-BRST symmetry transformation in the framework of the
superfield approach to the BRST formalism (cf. Subsec. 4.2 below).

In fact, we shall see that the cohomological origin of the {\it
co-existence} of the (anti-)co-BRST symmetry transformations {\it
together} for the 2D free Abelian gauge theory is encapsulated in
the existence of the super co-exterior derivative on the four (2,
2)-dimensional supermanifold which will be exploited in the
dual-horizontality condition.

We focus now on the existence of a bosonic (non-nilpotent)
symmetry transformation $s_w$ that emerges due to the
anticommutation relation between the nilpotent (anti-)BRST and
(anti-)co-BRST symmetry transformations (i.e. $s_w = s_b s_d + s_d
s_b \equiv s_{ab} s_{ad} + s_{ad} s_{ab}$). Under this symmetry
transformation, the fields of the Lagrangian density (2.2)
transform as follows (see, e.g. [36,38,41] for details) $$
\begin{array}{lcl}
 &&s_w A_\mu = \partial_\mu {\cal B} + \varepsilon_{\mu\nu}
\partial^\nu B,  \qquad s_w E = - \Box B, \qquad s_w (\partial \cdot
A) = \Box {\cal B}, \nonumber\\ &&s_w  C = 0, \qquad s_w \bar C =
0, \qquad s_w {\cal B} = 0, \qquad s_w B = 0.
\end{array}\eqno(2.5)
$$ It can be easily checked that the above transformations entail
upon the Lagrangian density (2.2) to change to a total derivative
as: $s_w {\cal L}_B = \partial_\mu \;[B \partial^\mu {\cal B} -
{\cal B} \partial^\mu B]$. The algebra followed by the above
transformation operators $s_r$ (with $ r = b, ab, d, ad, w$) is $$
\begin{array}{lcl}
&& s_{(a)b}^2 = 0, \qquad s_{(a)d}^2 = 0, \qquad s_w^2 \neq 0,
\nonumber\\ && s_w = \{s_b, s_d \} \equiv \{ s_{ab}, s_{ad} \},
\qquad \{s_b, s_{ab} \} = 0, \qquad \{s_d, s_{ad} \} = 0,
\nonumber\\ && \{s_b, s_{ad} \} = 0, \qquad \{s_d, s_{ab} \} = 0,
\qquad [s_w, s_{(a)b}] = [s_w, s_{(a)d}] = 0.
\end{array}\eqno(2.6)
$$ This algebra can be compared and contrasted with the algebra
obeyed by the de Rham cohomological operators as given below $$
\begin{array}{lcl}
&&d^2 = 0, \qquad \delta^2 = 0, \qquad \Delta = \{d, \delta \}
\equiv (d + \delta)^2, \nonumber\\ && [\Delta, d] = 0, \qquad
[\Delta, \delta] = 0, \qquad \{d , \delta \} \neq 0.
\end{array}\eqno(2.7)
$$ A close look at equations (2.6) and (2.7) establishes a
two-to-one correspondence between the symmetry transformation
operators $s_r$ (with $r = b, d, ab, ad, w$) and the cohomological
operators ($d, \delta, \Delta$). These are: $(s_b, s_{ad}) \to d,
(s_d, s_{ab}) \to \delta, \{s_b, s_d\} \equiv \{s_{ad}, s_{ab}\}
\to \Delta$. All the above continuous symmetry transformations are
generated by the conserved charges $Q_r$ ($r = b, ab, d, ad, w$)
and their intimate relationship can be succinctly expressed as $$
\begin{array}{lcl}
s_r \Omega = -\; i \;[\; \Omega,\; Q_r \;]_{\pm}, \;\qquad \;
\Omega = A_\mu, C, \bar C, B, {\cal B},
\end{array}\eqno(2.8)
$$ where the $(+)-$ signs on the square bracket stand for the
bracket to be an (anti)commutator for the generic field $\Omega$
of the Lagrangian density (2.2) being (fermionic)bosonic in
nature. It should be noted, at this stage, that the conserved
charges $Q_r$ (with $r = b, ab, d, ad, w$) obey exactly the same
kind of algebra as the one (cf. (2.6)) obeyed by the corresponding
symmetry transformation operators $s_r$. Furthermore, the mapping
between the conserved charges and the de Rham cohomological
operators are found to be exactly the same (i.e. $(Q_b, Q_{ad})
\to d, (Q_d, Q_{ab}) \to \delta$ and $Q_w = \{Q_b, Q_d \} =
\{Q_{ad}, Q_{ab} \} \to \Delta = \{d, \delta \}$).

It should be emphasized, as a side remark, that there is no effect
of the Laplacian operator $\Delta$ on any of the terms of the
Lagrangian density (2.2) because $\Delta A^{(1)} = dx^\mu \Box
A_\mu = 0$ leads to the derivation of the equation of motion
($\Box A_\mu = 0$) for the gauge field $A_\mu$ which (i.e. the
equation of motion) is {\it not} present in the Lagrangian density
(2.2) on its own. It emerges, however, from (2.2) due to the
application of the Euler-Lagrange equation of motion on it. This
observation will be exploited in Subsec. 4.3  where we shall show
that the appropriate definition of a super Laplacian operator, in
a suitable restriction on the four (2, 2)-dimensional
supermanifold, leads to the derivation of the equation of motion
for all the fields of the Lagrangian density (2.2) within the
framework of the superfield approach to BRST formalism. In fact,
the bosonic symmetry (cf. (2.5)) transformations $s_w = \{s_b, s_d
\} \equiv \{s_{ad}, s_{ab} \}$ encompass the analogue of the
definition of the Laplacian operator in terms of the cohomological
operators $d$ and $\delta$ as: $\Delta = \{d, \delta \} \equiv d
\delta + \delta d$.

Before we wrap up this Sect., we shall dwell a bit on the
existence of the discrete symmetry transformations in the theory.
First of all, it can be noted that the (anti-)ghost part of the
action (i.e. $S_{F.P} = - i \int d^2 x \partial_\mu \bar C
\partial^\mu C$) remains invariant under the following
discrete symmetry transformations (for the 2D 1-form Abelian gauge
theory): $$
\begin{array}{lcl}
C \to \pm \;i \;\bar C, \qquad \bar C \to \pm\; i\; C, \qquad
\partial_\mu \to \pm\; i\; \varepsilon_{\mu\nu}\; \partial^\nu.
\end{array}\eqno(2.9)
$$ The existence of the above discrete symmetry transformations is
responsible for the derivation of the (anti-)co-BRST symmetry
transformations from the (anti-)BRST symmetry transformations
[36,38,41]. It is already well known that the ghost action is also
invariant under $C \to \pm i \bar C, \bar C \to \pm i C$ which
leads to the derivation of (i) the anti-BRST symmetry
transformations from the BRST symmetry transformations, and (ii)
the anti-co-BRST symmetry transformations from the co-BRST
symmetry transformations. It can be checked explicitly that the
Lagrangian density (2.2) remains invariant under the following
separate and independent discrete symmetry transformations $$
\begin{array}{lcl}
&&C \to \pm \;i \;\bar C, \qquad \bar C \to \pm\; i\; C, \qquad
\partial_\mu \to \pm\; i\; \varepsilon_{\mu\nu}\; \partial^\nu
\qquad A_\mu \to A_\mu, \nonumber\\ && B \to \mp \;i \;{\cal B},
\quad {\cal B} \to \mp \;i \;B, \quad (\partial \cdot A) \to \pm
\;i\; E, \quad E \to \pm\; i\; (\partial \cdot A),
\end{array}\eqno(2.10)
$$
$$
\begin{array}{lcl}
&&C \to \pm \;i \;\bar C, \qquad \bar C \to \pm\; i\; C, \qquad
A_\mu \to \mp\; i\; \varepsilon_{\mu\nu}\; A^\nu \qquad
\partial_\mu \to \partial_\mu, \nonumber\\
&& B \to \mp \;i \;{\cal B}, \quad  {\cal B} \to \mp\; i\; B,
\quad (\partial \cdot A) \to \pm \;i\; E, \quad E \to \pm \;i
\;(\partial \cdot A).
\end{array}\eqno(2.11)
$$ The above transformations are found to be the analogue of the
Hodge duality $*$ operation of the differential geometry for the
1-form free Abelian gauge theory when they are combined with the
continuous and nilpotent (anti-)BRST and (anti-)co-BRST symmetry
transformations.

To elaborate a bit on the importance of the above discrete
symmetry transformations, we, first of all, check the effect of
two successive discrete transformations (2.11) on any arbitrary
generic field $\Omega$ of the Lagarangian density (2.2). This is
required as an essential ingredient for the discussion of any
arbitrary {\it duality} invariant theory [48]. In fact, this
requirement decides the relationship between a physical quantity
and its dual. For the theory under consideration, this requirement
is given by the following expression $$
\begin{array}{lcl}
*\; (*\; \Omega) = - \; \Omega, \qquad \Omega = C, \bar C, A_\mu,
B, {\cal B}, E, (\partial \cdot A).
\end{array}\eqno(2.12)
$$ The above $(-)$ sign on the r.h.s. dictates the following
interesting relationship $$
\begin{array}{lcl}
s_{(a)d}\; (\Omega) = - \; *\; s_{(a)b} \;* \;(\Omega),
\end{array}\eqno(2.13)
$$ which is a relationship that involves the interplay between the
discrete symmetry transformation (2.11) (or (2.10)) and continuous
symmetry transformations $s_{(a)b}$ and $s_{(a)d}$. It is evident
that the above relationship is (i) the analogue of the
relationship that exists between the exterior derivative $d =
dx^\mu \partial_\mu$ and the dual(co)-exterior derivative (i.e.
$\delta = - * d *$), and (ii) contains the definition of the Hodge
duality $*$ operation as the discrete symmetry transformations
(2.11) (or (2.10)). In more transparent language, the operation of
the cohomological differential operators $\delta$ and $d$ on a
differential form is equivalent to the operation of the nilpotent
symmetry transformations $s_{(a)d}$ and $s_{(a)b}$ on the generic
field $\Omega$ of the Lagrangian density (2.2) of the 2D 1-form
free Abelian gauge theory. This statement has been mathematically
expressed in (2.13). It is worthwhile to mention, at this stage,
that there is a reciprocal relationship of the above equation,
namely; $$
\begin{array}{lcl}
s_{(a)b}\; (\Omega) = - \; *\; s_{(a)d} \;* \;(\Omega),
\end{array}\eqno(2.14)
$$ that also exists on the 2D spacetime manifold. In fact, the
duality operation on the 4D and 2D spacetime manifolds are
different as can be seen in [48]. For more discussions on the
relationships between the Hodge duality $*$ operation and discrete
symmetry transformations of the theory, we refer to the readers
our earlier works on this subject [41,38,36].\\

\noindent {\bf 3. Topological Features of 2D 1-Form Abelian Gauge
Theory}\\

\noindent In this Sec., first of all, we express the form of the
Lagrangian density (2.1) in terms of the conserved and on-shell
nilpotent (anti-)BRST and (anti-)co-BRST charges. In Subsec. 3.1,
we focus, for our elaborate discussions, only on the Lagrangian
density (2.1) because it contains merely the basic dynamical
fields of the theory (i.e. there is {\it no} presence of any
auxiliary fields in it). Our Subsec. 3.2 is devoted to the
discussion of the BRST cohomology, physical state condition and
topological nature of the theory in terms of the BRST and co-BRST
charges and their action on the states of the quantum Hilbert
space.\\

\noindent {\bf 3.1 Topological Aspects: Mathematical
Description}\\

\noindent In this Subsec., we demonstrate that the form of the
Lagrangian density is similar to the Witten-type topological field
theory. To corroborate the above assertion, let us concentrate on
the Lagrangian density (2.1) and express it as follows $$
\begin{array}{lcl}
{\cal L}_b = \tilde s_b\; (i T_1) + \tilde s_d \;(i T_2) +
\;\partial_\mu Y^\mu \equiv \tilde s_{ab}\; (i P_1) + \;\tilde
s_{ad}\; (i P_2) + \;\partial_\mu Y^\mu.
\end{array}\eqno(3.1)
$$ The following points, at this stage, are in order now. First,
in the above equation, $\tilde s_{(a)b}$ and $\tilde s_{(a)d}$ are
the on-shell ($\Box C = 0, \Box \bar C = 0$) nilpotent ($\tilde
s_{(a)b}^2 = 0, \tilde s_{(a)d}^2 = 0$) version of the (anti-)BRST
and (anti-)co-BRST symmetry transformations. Explicitly these are
$$
\begin{array}{lcl}
&&\tilde s_b A_\mu = \partial_\mu C, \qquad \tilde s_b C = 0,
\qquad \tilde s_b \bar C = - i (\partial \cdot A), \nonumber\\ &&
\tilde s_b E = 0, \qquad \tilde s_b (\partial \cdot A) = \Box C,
\qquad \tilde s_b F_{\mu\nu} = 0, \nonumber\\ && \tilde s_{ab}
A_\mu = \partial_\mu \bar C, \qquad \tilde s_{ab} \bar C = 0,
\qquad \tilde s_{ab} C = + i (\partial \cdot A), \nonumber\\ &&
\tilde s_{ab} E = 0, \qquad \tilde s_{ab} (\partial \cdot A) =
\Box \bar C, \qquad \tilde s_{ab} F_{\mu\nu} = 0,
\end{array} \eqno(3.2)
$$ $$
\begin{array}{lcl}
 &&\tilde s_d A_\mu = - \varepsilon_{\mu\nu}
\partial^\nu \bar C, \qquad \tilde s_d \bar C = 0,
\qquad \tilde s_d  C = - i E, \nonumber\\ && \tilde s_d E = \Box
\bar C, \qquad \tilde s_d (\partial \cdot A) = 0, \qquad \tilde
s_d F_{\mu\nu} = [\varepsilon_{\mu\rho} \partial_\nu -
\varepsilon_{\nu\rho}
\partial_\mu]\; \partial^\rho \bar C, \nonumber\\
&&\tilde s_{ad} A_\mu = - \varepsilon_{\mu\nu}
\partial^\nu  C, \qquad \tilde s_{ad}  C = 0, \qquad
\tilde s_{ad} \bar C =
 i E, \nonumber\\ && \tilde s_{ad} E
= \Box C, \qquad \tilde s_{ad} (\partial \cdot A) = 0, \qquad
\tilde s_{ad} F_{\mu\nu} = [\varepsilon_{\mu\rho} \partial_\nu -
\varepsilon_{\nu\rho}
\partial_\mu]\; \partial^\rho  C.
\end{array}\eqno(3.3)
$$ The above equations (3.2) and (3.3) have been derived from
(2.3) and (2.4) by the substitution $B = - (\partial \cdot A)$ and
${\cal B} = E$. Second, the exact expressions for $T_1, T_2, P_1,
P_2$ and $Y^\mu$, for the Lagrangian density (2.1), are as follows
$$
\begin{array}{lcl}
&&{\displaystyle T_1 = - \frac{1}{2}\; (\partial \cdot A)\; \bar
C, \quad T_2 = + \frac{1}{2}\; E\; C, \quad P_1 = + \frac{1}{2}
(\partial \cdot A)\; C, \quad P_2 = - \frac{1}{2}\; E \;\bar C,}
\nonumber\\ && {\displaystyle Y^\mu = \frac{i}{2}\; [\;\bar C
\partial^\mu C\; + \;\partial^\mu \bar C C \;] \equiv \;\frac{i}{2}\;
[\;\partial^\mu (\bar C \; C)\;]}.
\end{array}\eqno(3.4)
$$ Finally, taking the help of equation (2.8), it can be seen that
the Lagrangian density (2.1), modulo some total derivatives, can
also be written as the sum of two anticommutators, namely;$$
\begin{array}{lcl}
{\displaystyle {\cal L}_b = \{ \tilde Q_b, T_1 \}+ \{\tilde Q_d,
T_2 \} \equiv \{ \tilde Q_{ab},  P_1 \} + \{ \tilde Q_{ad},  P_2
\}},
\end{array}\eqno(3.5)
$$ where $\tilde Q_r$ (with $r = b, ab, d, ad$) are the on-shell
nilpotent version of the conserved and nilpotent (anti-)BRST and
(anti-)co-BRST charges. The exact expressions for these charges
are not essential for us at the moment. However, their exact
expressions, in terms of the local fields of the theory, can be
found in [46,47] and in Subsec. 3.2 (cf. (3.15)).

At this juncture, a few further remarks are in order. First, even
though we have two pairs of nilpotent charges (i.e. $(\tilde Q_b,
\tilde Q_d)$ as well as $(\tilde Q_{ad}, \tilde Q_{ab})$) in terms
of which the Lagrangian density of the theory can be expressed,
still we claim that the Lagrangian density looks like Witten-type
topological theory. The key reason behind this assertion is the
fact that this field theoretic model is a tractable model for the
Hodge theory. As a consequence, we can choose the physical state
$|phys>$ to be the harmonic state in the Hodge decomposition
theorem (see, Subsec. 3.2 for details) because this state happens
to be the most symmetric state. In fact, the physical state,
chosen in such a manner, is (anti-)BRST invariant (i.e. $\tilde
Q_{(a)b} |phys> = 0$) as well as (anti-)co-BRST invariant (i.e.
$\tilde Q_{(a)d} |phys> = 0$) {\it together}. Precisely speaking,
for the Witten type topological field theory, there exist only
nilpotent (anti-)BRST charges and the physical states are
annihilated by these (i.e. $\tilde Q_{(a)b} |phys> = 0$) {\it
alone}. Second, it is evident that the form of the Lagrangian
density (3.1) and (3.5) is not like the Schwarz type topological
field theory (TFT) because, for such kind of a TFT, there exists
always a piece in the Lagrangian density that can not be expressed
as a BRST (anti)commutator. Third, the symmetry transformations
for the free 2D 1-form Abelian gauge theory is like the Schwarz
type TFT because there exists no local topological shift symmetry
for the 1-form Abelian gauge theory (which happens to be the
hallmark for a Witten type TFT). Thus, from the symmetry point of
view, our prototype field theoretical model of 2D 1-form gauge
theory is like the Schwarz type TFT. Finally, we are considering
our present 2D 1-form gauge model in the flat Minkowskian
spacetime. Thus, there is no non-trivial metric dependence in the
theory. As a consequence, the model of our present discussion,
once again, is like the Schwarz type TFT. Thus, finally, we
conclude that the form of the Lagrangian density of the 2D 1-form
gauge theory is like the Witten type TFT but the symmetries of the
theory are like the Schwarz type of TFT.

One of the decisive features of the TFTs is the absence of any
energy excitations in the theory which is mainly governed and
dictated by the form of the symmetric energy-momentum tensor
($T_{\mu\nu}^{(s)} = T_{\nu\mu}^{(s)}$) of the theory. This
expression for the case of the free 1-form Abelian gauge theory,
described by the Lagrangian density (2.1), is
 $$
\begin{array}{lcl}
{\displaystyle  T_{\mu\nu}^{(s)} = \frac{1}{2}\; \partial_\mu
\Omega\; \frac{\partial {\cal L}_b}{\partial_\nu \Omega} +
\frac{1}{2}\;
\partial_\nu \Omega\; \frac{\partial {\cal L}_b}{\partial_\mu
\Omega} - \eta_{\mu\nu}\; {\cal L}_b},
\end{array}\eqno(3.6)
$$ where the generic field $\Omega = A_\mu, C, \bar C$ for the
Lagrangian density (2.1). The exact and explicit expression for
the above symmetric tensor is $$
\begin{array}{lcl}
T_{\mu\nu}^{(s)} &=& - {\displaystyle \frac{1}{2} \bigl [ \;
\varepsilon_{\mu\rho} E + \eta_{\mu\rho} \;(\partial \cdot A)
\;\bigr ]\; (\partial_\nu A^\rho) - \frac{1}{2} \bigl [\;
\varepsilon_{\nu\rho} E + \eta_{\nu\rho}\; (\partial \cdot A)\;
\bigr ]\; (\partial_\mu A^\rho)} \nonumber\\ &-& i \partial_\mu
\bar C \partial_\nu C - i \partial_\nu \bar C
\partial_\mu C - \eta_{\mu\nu} \; {\cal L}_b.
\end{array}\eqno(3.7)
$$ The above expression can be expressed, in terms of the on-shell
nilpotent (anti-)BRST charge $\tilde Q_{(a)b}$ and (anti-)co-BRST
charge $\tilde Q_{(a)d}$, as $$
\begin{array}{lcl}
T_{\mu\nu}^{(s)} &=& \{\tilde Q_b, V_{\mu\nu}^{(1)} \} + \{\tilde
Q_d, V_{\mu\nu}^{(2)} \} \equiv \{\tilde Q_{ab}, \bar
V_{\mu\nu}^{(1)} \} + \{\tilde Q_{ad}, \bar V_{\mu\nu}^{(2)} \},
\nonumber\\ V_{\mu\nu}^{(1)} &=& + \frac{1}{2}\; \bigl [
(\partial_\mu \bar C)\; A_\nu + (\partial_\nu \bar C)\; A_\mu +
\eta_{\mu\nu} (\partial \cdot A)\; \bar C \bigr ], \nonumber\\
V_{\mu\nu}^{(2)} &=& + \frac{1}{2}\; \bigl [ (\partial_\mu C)\;
\varepsilon_{\nu\rho} A^\rho + (\partial_\nu C)\;
\varepsilon_{\mu\rho}\;A^\rho - \eta_{\mu\nu}\; E\; C \bigr
],\nonumber\\ \bar V_{\mu\nu}^{(1)} &=& - \frac{1}{2}\; \bigl [
(\partial_\mu C)\; A_\nu + (\partial_\nu C)\; A_\mu +
\eta_{\mu\nu} (\partial \cdot A)\;  C \bigr ], \nonumber\\ \bar
V_{\mu\nu}^{(2)} &=& - \frac{1}{2}\; \bigl [ (\partial_\mu \bar
C)\; \varepsilon_{\nu\rho} A^\rho + (\partial_\nu \bar C)\;
\varepsilon_{\mu\rho}\;A^\rho - \eta_{\mu\nu}\; E\; \bar C \bigr
].
\end{array}\eqno(3.8)
$$ The form of the symmetric energy-momentum tensor
($T_{\mu\nu}^{(s)}$) demonstrates that, when the Hamiltonian
density $T_{00}^{(s)}$ is sandwiched between two physical states,
it becomes zero because of the fact that $\tilde Q_{(a)b} |phys> =
0$ and $\tilde Q_{(a)d} |phys> = 0$ when we choose the physical
state to be the harmonic state of the Hodge decomposed state in
the quantum Hilbert space of states (cf. Subsec. 3.2). In fact,
this state is annihilated by all the conserved, {\it hermitian}
and nilpotent charges of the theory (i.e. $\tilde Q_{(a)b} |phys>
= 0, \tilde Q_{(a)d} |phys> = 0$). To be precise, the conditions
$\tilde Q_b |phys> = 0$ and $\tilde Q_d |phys> = 0$ lead to the
restrictions $(\partial \cdot A) |phys> = 0$ and
$\varepsilon^{\mu\nu}
\partial_\mu A_\nu |phys> = 0$, respectively. This ensures that
there are no propagating degrees of freedom in the theory as both
the components $A_0$ and $A_1$ of the 2D gauge field become
conserved quantities (i.e. $\partial_0 A_0 = \partial_1 A_1,
\partial_0 A_1 = \partial_1 A_0$). This situation
is the most remarkable feature of a TFT. Thus, the topological
nature of the 2D free 1-form Abelian gauge theory is confirmed
because there are no propagating dynamical degrees of freedom left
out with the 2D gauge field $A_\mu$, after the application of the
physicality criteria (i.e. $\tilde Q_{(a)b} |phys> = 0, \tilde
Q_{(a)d} |phys> = 0$). The present theory is, however, a {\it new}
kind of TFT which captures together some of the salient features
of {\it both} the Witten-type as well as the Schwarz-type TFTs.
This key observation is a completely new result [41,36].\\

\noindent {\bf 3.2 Cohomological Aspects: Topological Features}\\

\noindent In this Subsec., we shall discuss the Hodge
decomposition theorem and establish the topological nature of the
theory due to the consideration of the cohomological aspects of
the states in the quantum Hilbert space. To this end in mind, let
us begin with the infinitesimal version of the ghost symmetry
transformations $s_g$ for the Lagrangian density (2.1), namely;$$
\begin{array}{lcl}
s_g A_\mu  = 0, \qquad  s_g C = - \Sigma \;C, \qquad s_g \bar C =
+ \Sigma \;\bar C,
\end{array}\eqno(3.9)
$$ where $\Sigma$ is a global scale transformation parameter.
Under the above symmetry transformation, the Lagrangian density
(2.1) remains invariant. The generator (i.e. ghost charge) $\tilde
Q_g$ of the above transformations is a conserved quantity which
obeys the following algebra with the rest of the generators
$\tilde Q_r$ (with $r = b, ab, d, ad, w$) of the theory (see, e.g.
[36]) $$
\begin{array}{lcl}
&&[\tilde Q_w, \tilde Q_g ] = 0, \qquad \; \;i [\tilde Q_g, \tilde
Q_b] = +\; \tilde Q_b, \qquad \;\;\;i [\tilde Q_g, \tilde Q_d ] =
- \;\tilde Q_d, \nonumber\\ && i [\tilde Q_g, \tilde Q_{ab}] = -
\;\tilde Q_{ab}, \;\;\;\qquad \;\;\;i [\tilde Q_g, \tilde Q_{ad} ]
= + \;\tilde Q_{ad}.
\end{array}\eqno(3.10)
$$ The above algebra plays a very important role in the discussion
of the Hodge decomposition theorem. This is due to the fact that,
given any arbitrary state $|\phi>_n$ with the ghost number n (i.e.
$i \tilde Q_g |\phi>_n = n \; |\phi>_n $) in the quantum Hilbert
space of states, it is clear, from the above algebra, that the
following relationships are true, namely;$$
\begin{array}{lcl}
 i\; \tilde Q_g\; \tilde Q_b \; |\phi>_n &=& (n + 1)\; \tilde Q_b\;
 |\phi>_n, \qquad
 i\; \tilde Q_g\; \tilde Q_{ab} \; |\phi>_n = (n - 1)\; \tilde Q_{ab}\;
 |\phi>_n, \nonumber\\ i\; \tilde Q_g\; \tilde Q_d \; |\phi>_n
 &=& (n - 1)\; \tilde Q_d\;
 |\phi>_n, \qquad
i\;\tilde Q_g\; \tilde Q_{ad} \; |\phi>_n = (n + 1)\; \tilde
Q_{ad}\;
 |\phi>_n, \nonumber\\
 i\; \tilde Q_g\; \tilde Q_w \; |\phi>_n\;
 & = & n\; \tilde Q_w\; |\phi>_n.
\end{array}\eqno(3.11)
$$ The above equation shows that the ghost number of the states
$\tilde Q_b |\phi>_n$ and $\tilde Q_{ad} |\phi>_n$ is $(n + 1)$.
This situation is like $d f_n \sim \tilde f_{n + 1}$ in the
differential geometry (when the exterior derivative $d = dx^\mu
\partial_\mu$ acts on the n-form $f_n$). On the contrary, the
states $\tilde Q_d |\phi>_n$ and $\tilde Q_{ab} |\phi>_n$ carry
the ghost number equal to $(n - 1)$. This feature is like $\delta
f_n \sim \hat f_{n-1}$ due to the operation of the co-exterior
derivative $\delta$ on the n-form $f_n$. In view of the forms of
the algebra illustrated in (2.6) and (2.7) for the symmetry
operators and the cohomological operators, it is clear that one
can have the analogue of the Hodge decomposition theorem (1.1) in
the Hilbert space of states as given below (see, e.g. [41,42] for
details) $$
\begin{array}{lcl}
 |\phi>_n &=& |\omega>_{(n)} + \tilde Q_b \; | \theta>_{(n - 1)} + \tilde
 Q_d \;
|\chi>_{(n + 1)}, \nonumber\\&  \equiv & |\omega>_{(n)} + \tilde
Q_{ad} \; | \theta>_{(n - 1)} + \tilde Q_{ab} \; |\chi>_{(n + 1)},
\end{array}\eqno(3.12)
$$ where $\tilde Q_b |\theta>_{(n - 1)}$ is the BRST exact state
and $\tilde Q_d |\chi>_{(n + 1)}$ is the BRST co-exact state.
There are two ways to write the Hodge decomposition theorem in the
quantum Hilbert space of states because there is two-to-one
mapping between the conserved charges and the cohomological
operators (i.e. $(\tilde Q_b, \tilde Q_{ad}) \to d, (\tilde Q_d,
\tilde Q_{ab}) \to \delta, \{\tilde Q_b, \tilde Q_d \} = \{\tilde
Q_{ad}, \tilde Q_{ab} \} \to \Delta$). In the above decomposition,
the most symmetric state $|\omega>_{(n)}$ is the harmonic state
that is annihilated (i.e. $\tilde Q_r \;|\omega>_{(n)} = 0$) by
all the charges $\tilde Q_r$ (with $r = b, ab, d, ad, w$). In
other words, the harmonic state is {\it already} an (anti-)BRST
and (anti-)co-BRST invariant state, to begin with. This is why, it
is the most beautiful state of our present 2D free Abelian theory.
It will be chosen, therefore, as the physical state on an
aesthetic ground.

To observe the key consequences of the above statements, let us
begin with our physical state to be the harmonic state of the
above decomposition (i.e. $|phys> = |\omega>$). This immediately
implies the following $$
\begin{array}{lcl}
\tilde Q_b \; |phys> = 0, \qquad \tilde Q_d \; |phys> = 0, \qquad
\tilde Q_w \; |phys> = 0.
\end{array}\eqno(3.13)
$$ It will be noted that, in the above, we have taken only one set
(i.e. $\tilde Q_b, \tilde Q_d, \tilde Q_w$) of the conserved
charges, as the consequences from this set, would be exactly the
same as the ones derived from the other set ($\tilde Q_{ad},
\tilde Q_{ab}, \tilde Q_w$). The basic conditions, that emerge
from the above restrictions, are as follows $$
\begin{array}{lcl}
&& \tilde Q_b \; |phys> = 0\;\; \Rightarrow\;\; (\partial \cdot A)
|phys> = 0, \qquad \partial_0 (\partial \cdot A) |phys> = 0,
\nonumber\\ && \tilde Q_d \; |phys> = 0 \;\;\;\Rightarrow
\;\;\;E\; |phys> = 0, \;\;\;\qquad \;\;\;\partial_0 E\; |phys> =
0,
\end{array}\eqno(3.14)
$$ where the explicit expressions for the charges $\tilde Q_b ,
\tilde Q_d, \tilde Q_w$ are $$
\begin{array}{lcl}
&&\tilde Q_b = {\displaystyle \int} \;(dx) \;\bigl [\; \partial_0
(\partial \cdot A)\; C - (\partial \cdot A)\; \dot C \;\bigr ],
\qquad \tilde Q_d = {\displaystyle \int}\; (dx) \;\bigl [\; E\;
\dot {\bar C} - \dot E\; \bar C\; \bigr ], \nonumber\\ &&\tilde
Q_w = {\displaystyle \int}\; (dx) \;\bigl [\; \partial_0 (\partial
\cdot A)\; E - \dot E\; (\partial \cdot A)\; \bigr ].
\end{array}\eqno(3.15)
$$ The conditions generated by $\tilde Q_w$ are not new. They are
same as the ones emerging due to $\tilde Q_b$ and $\tilde Q_d$. We
shall dwell a bit more on the above conditions (3.14) in the
language of the normal mode expansions of the basic fields of the
Lagrangian density (2.1). For this purpose, let us have the normal
mode expansion (for the equations of motion $\Box A_\mu = 0, \Box
C = 0$ and $\Box \bar C = 0 $) in the phase space of the theory,
as (see, e.g., [47]) $$
\begin{array}{lcl}
A_\mu (x, t) &=& {\displaystyle \int\; \frac{dk}{(2\pi)^{1/2} (2
k_0)^{1/2}}} \; \bigl [\; a_\mu (k) e^{-i k \cdot x} +
a_\mu^{\dagger} (k) e^{+ i k \cdot x} \;\bigr ], \nonumber\\ C (x,
t) &=& {\displaystyle \int\; \frac{dk}{(2\pi)^{1/2} (2
k_0)^{1/2}}} \; \bigl [\; c (k) e^{-i k \cdot x} + c^{\dagger} (k)
e^{+ i k \cdot x}\; \bigr ], \nonumber\\ \bar C (x, t) &=&
{\displaystyle \int\; \frac{dk}{(2\pi)^{1/2} (2 k_0)^{1/2}}} \;
\bigl [\; b(k) e^{-i k \cdot x} +  b^{\dagger} (k) e^{+ i k \cdot
x} \;\bigr ],
\end{array}\eqno(3.16)
$$ where $k_\mu = (k_0, k_1 = k)$ is the 2D momentum vector in the
phase space and $a_\mu^{\dagger}, c^\dagger$ and $b^\dagger$ are
the creation operators for a photon, a ghost quantum and an
anti-ghost quantum, respectively. The corresponding operators,
without a dagger, are the annihilation operators.

Taking the help of the nilpotent symmetry transformations of (3.2)
and (3.3), we can obtain the following (anti)commutators $$
\begin{array}{lcl}
&& [\tilde Q_b, a_\mu^\dagger] = + k_\mu c^\dagger (k), \qquad
\;\;\;[\tilde Q_d, a_\mu^\dagger ] = - \varepsilon_{\mu\nu} k^\nu
b^\dagger (k), \nonumber\\ && [\tilde Q_b, a_\mu ] = - k_\mu c
(k), \;\;\qquad\;\;\; [\tilde Q_d, a_\mu ] = +
\varepsilon_{\mu\nu} k^\nu b (k), \nonumber\\&& \{ \tilde Q_b,
c^\dagger (k) \} = 0, \;\;\;\;\;\qquad\;\;\;\;\; \{ \tilde Q_d,
c^\dagger (k) \} = - i \varepsilon^{\mu\nu} k_\mu a^{\dagger}_\nu,
\nonumber\\&& \{ \tilde Q_b, c (k) \} = 0,
\;\;\;\qquad\;\;\;\;\;\;\;\;\; \{\tilde Q_d, c (k) \} = + i
\varepsilon^{\mu\nu} k_\mu a_\nu, \nonumber\\ &&\{ \tilde Q_b,
b^\dagger (k) \} = + i k^\mu a_\mu^{\dagger}, \;\qquad \{\tilde
Q_d, b^\dagger (k) \} = 0, \nonumber\\ &&\{ \tilde Q_b, b (k) \} =
- i k^\mu a_\mu, \;\qquad\;\; \{\tilde Q_d, b (k) \} = 0,
\end{array}\eqno(3.17)
$$ where (i) we have exploited the expansion given in (3.16), and
(ii) we have utilized the formula, given in (2.8), between the
infinitesimal nilpotent transformations $\tilde s_r$ (with $r = b
, d$) and their generators $\tilde Q_r$ (with $r = b, d$). It is
evident, from the expression for $\tilde Q_w$ in (3.15), that this
conserved charge generates the following transformations $$
\begin{array}{lcl}
&&\tilde s_w A_\mu = \partial_\mu E - \varepsilon_{\mu\nu}
\partial^\nu (\partial \cdot A), \qquad \;\;\;\tilde s_w E = + \Box
(\partial \cdot A), \nonumber\\&& \tilde s_w (\partial \cdot A) =
\Box E, \qquad \tilde s_w C = 0, \qquad \tilde s_w \bar C = 0,
\end{array}\eqno(3.18)
$$ which, ultimately, leads to the following commutation relations
$$
\begin{array}{lcl}
&& [Q_w, a_\mu^\dagger (k)] = + i\;k^2\;\varepsilon_{\mu\nu}
(a^\nu (k))^\dagger, \qquad  [Q_w, a_\mu (k) ] = - i\; k^2\;
\varepsilon_{\mu\nu} a^\nu (k), \nonumber\\ && [Q_w, c(k)] = [Q_w,
c^\dagger (k)] = [ Q_w, b (k) ] = [Q_w, b^\dagger (k) ] = 0,
\end{array}\eqno(3.19)
$$ where, once again, the mode expansion of (3.16) and analogue of
the relation (2.8) for $\tilde s_w$ and $\tilde Q_w$ (i.e. $\tilde
s_w \Omega = - i [\Omega, \tilde Q_w]$) have been used. For the
masslessness condition (i.e. $k^2 = 0$), it is evident that the
charge $\tilde Q_w$ will become the Casimir operator.

Let us now define the physical vacuum of the theory by the
following conditions that are imposed by its {\it physical}
properties and its {\it harmonic} nature, namely; $$
\begin{array}{lcl}
&&\tilde Q_b \;|vac> = 0, \qquad \tilde Q_d \;|vac> = 0, \qquad
\tilde Q_w\; |vac> = 0, \nonumber\\ && a_\mu (k)\; |vac> = 0,
\qquad c (k)\; |vac> = 0, \qquad b (k) \; |vac> = 0.
\end{array}\eqno(3.20)
$$ From the above vacuum state, a single {\it physical} photon
state with the polarization $e_\mu$ can be created by the
application of the creation operator $a_\mu^\dagger$. This can be
expressed as: $e^\mu a_{\mu}^\dagger\; |vac> \equiv |e, vac>$
(see, e.g. [47]). In an exactly similar fashion, a single photon
with momentum $k^\mu$ can be created from the vacuum state by
$k^\mu a_{\mu}^\dagger \;|vac> \equiv |k, vac>$. The single
physical photon state being a harmonic state, the following
restrictions emerge due to the application of the conserved
charges $\tilde Q_r$ (with $ r = b, d, w$) on it, namely; $$
\begin{array}{lcl}
&&\tilde Q_b \;|e, vac> \equiv [\tilde Q_b, e^\mu a_\mu^\dagger]
\; |vac> = 0 \Rightarrow (k\cdot e) = 0, \nonumber\\ &&\tilde Q_d
\;|e, vac> \equiv [\tilde Q_d, e^\mu a_\mu^\dagger] \; |vac> = 0
\Rightarrow (\varepsilon_{\mu\nu} e^\mu k^\nu) = 0, \nonumber\\
&&\tilde Q_w \;|e, vac> \equiv [\tilde Q_w, e^\mu a_\mu^\dagger]
\;|vac> = 0 \Rightarrow (k^2) = 0.
\end{array}\eqno(3.21)
$$ The above relations are found to be consistent with
one-another. In fact, any two of the above relations imply the
third one [36]. In particular, it will be noted that the relations
(i.e $(k \cdot e) = 0, \varepsilon_{\mu\nu} e^\mu k^\nu = 0$)
emerging from the BRST charge $\tilde Q_b$ as well as co-BRST
charge $\tilde Q_d$ are invariant under the gauge transformation $
e_\mu \to e_\mu + \alpha\; k_\mu$ and the dual gauge
transformation $e_\mu \to e_\mu + \beta\;\varepsilon_{\mu\nu} \;
k^\nu$ if the masslessness condition $k^2 = 0$ is taken into
account. Here $\alpha$ and $\beta$, in the above (dual-)gauge
transformations, are spacetime independent constants. The above
two symmetry transformations are good enough to gauge away {\it
both} the degrees of freedom of photon in two (1 + 1)-dimensions
of spacetime. This is the root cause of the 2D photon to be
topological in nature because there are no {\it propagating}
degrees of freedom left out in the free 2D 1-form Abelian gauge
theory [36].

We would like to close this subsection with the remark that the
existence of the BRST and co-BRST symmetry transformations enables
one to decompose both the degrees of freedom of the 2D photon
(i.e. $dx^\mu A_\mu (x) = dx^\mu \partial_\mu \kappa (x) + dx^\mu
\varepsilon_{\mu\nu} \partial^\nu \zeta (x)$) into (i) a component
parallel to the momentum vector, and (ii) the other component
parallel to the polarization vector. Here the fields $\kappa (x)$
and $\zeta (x)$ are the component fields. Thus, a 2D photon has no
propagating (dynamical) degrees of freedom. As a consequence, the
2D free 1-form Abelian gauge theory is a (new kind of) topological
field theory (see, e.g., [36] for details).\\

\noindent {\bf 4 Superfield Approach to BRST Formulation of 2D
Gauge Theory}\\

\noindent In this Sec., we shall tap the power and potential of
the super de Rham cohomological operators and demonstrate their
usefulness in the derivation of (i) the off-shell nilpotent
(anti-)BRST symmetries, (ii) the off-shell nilpotent
(anti-)co-BRST symmetries, and (iii) the equations of motion for
{\it all} the fields of the 2D free 1-form Abelian gauge theory.
Furthermore, we shall also capture the topological features of the
above theory in the language of the superfield approach to BRST
formalism.\\

\noindent
{\bf 4.1 Super Exterior Derivative and (Anti-)BRST Symmetries}\\

\noindent As pointed out after equation (2.3), we know that the
cohomological origin for the nilpotent (anti-)BRST symmetry
transformations lies in the nilpotent exterior derivative. This
statement becomes very clear in the framework of the superfield
approach to BRST formalism where we exploit the horizontality
condition (HC) for the derivation of the (anti-)BRST symmetry
transformations for the gauge and (anti-)ghost fields of the 2D
Abelian 1-form gauge theory. It turns out that the celebrated HC
[20-29], on the four (2, 2)-dimensional supermanifold, owes its
origin to the (super) exterior derivatives.

To elaborate a bit on the above assertion, we begin with the super
1-form connection $\tilde A^{(1)} = d Z^M A_M$ where (i) the
superspace variable on the four (2, 2)-dimensional supermanifold
is $Z^M = (x^\mu, \theta, \bar\theta)$ where $x^\mu$ (with $\mu =
0, 1$) are the 2D bosonic spacetime variable and $\theta,
\bar\theta$ are the Grassmannian variables (with $\theta^2 =
\bar\theta^2 = 0, \theta \bar\theta + \bar\theta \theta = 0$), and
(ii) the supermultiplet fields ${\cal B}_\mu
(x,\theta,\bar\theta), {\cal F} (x,\theta,\bar\theta), \bar {\cal
F} (x,\theta,\bar\theta)$ on the above supermanifold (as the
generalization of the 2D basic fields $A_\mu (x), C (x), \bar C
(x)$) constitute the vector superfield $A_M (x, \theta,
\bar\theta)$. Similarly, the exterior derivative $d = dx^\mu
\partial_\mu$ of the 2D Minkowskian spacetime manifold is
generalized to the super exterior derivative $\tilde d$ on the
four (2, 2)-dimensional supermanifold. The above statements can be
succinctly expressed, in the mathematical form, as: $$
\begin{array}{lcl}
\tilde d &=& d Z^M \partial_M = dx^\mu \;\partial_\mu + d \theta\;
\partial_\theta + d \bar\theta\; \partial_{\bar\theta}, \;\qquad\;
\tilde d^2 = 0, \nonumber\\ \tilde A^{(1)} &=& dZ^M A_M = dx^\mu
\;{\cal B}_\mu (x, \theta, \bar\theta)+ d \theta\; \bar {\cal F}
(x,\theta,\bar\theta) + d \bar\theta\; {\cal F}
(x,\theta,\bar\theta),
\end{array}\eqno(4.1)
$$
where $\partial_M = (\partial/ \partial Z^M) \equiv (\partial_\mu,
\partial_\theta, \partial_{\bar\theta})$ and  $A_M (x, \theta, \bar\theta)
= \bigl ({\cal B}_\mu (x,\theta,\bar\theta), {\cal F}
(x,\theta,\bar\theta), \bar {\cal F} (x,\theta,\bar\theta) \bigr )
$.

The above multiplet superfields can be expanded, along the
Grassmannian (i.e. $\theta, \bar\theta$) directions of the four
(2, 2)-dimensional supermanifold, as follows $$
\begin{array}{lcl}
{\cal B}_\mu (x, \theta, \bar\theta) &=& A_\mu (x) + \theta\; \bar
R_\mu (x) + \bar\theta \; R_\mu (x) + i\;\theta\; \bar\theta\;
S_\mu (x), \nonumber\\ {\cal F} (x, \theta, \bar\theta) &=& C (x)
+ i\;\theta\; \bar B_1 (x) + i\; \bar\theta \; B_1 (x) +
i\;\theta\; \bar\theta\; s (x), \nonumber\\ \bar {\cal F} (x,
\theta, \bar\theta) &=& \bar C (x) + i\;\theta\; \bar B_2 (x) +
i\;\bar\theta \; B_2 (x) + i\;\theta\; \bar\theta\; \bar s (x).
\end{array}\eqno(4.2)
$$ The points to be noted, at this juncture, are\\ (i) the above
expansion is achieved in terms of the basic fields $(A_\mu, C,
\bar C)$ and some secondary fields ($R_\mu, \bar R_\mu, S_\mu,
B_1, \bar B_1, B_2, \bar B_2, s , \bar s$) which are all functions
of the 2D spacetime variable $x^\mu$ alone as they are local
fields on the above manifold,\\ (ii) in the limit $(\theta \to 0,
\bar\theta \to 0)$, (a) we retrieve the 2D local basic fields
$(A_\mu, C, \bar C)$ of the Lagrangian density (2.2), and (b) the
super exterior derivative $\tilde d$ reduces to the 2D ordinary
exterior derivative $d = dx^\mu
\partial_\mu$, and\\ (iii) the number of the fermionic component
fields $(R_\mu, \bar R_\mu, C, \bar C, s, \bar s)$  do match with
the number of the bosonic fields $(A_\mu, S_\mu, B_1, \bar B_1,
B_2, \bar B_2)$ in the above expansion.

Now we are all set to exploit the celebrated horizontality
condition [20-29] which leads to (i) the exact expression for the
secondary fields in terms of the basic fields of the Lagrangian
density (2.2), and (ii) the derivation of the off-shell nilpotent
\footnote{The on-shell nilpotent symmetry transformations have
been derived in [49-51] where the (anti-)chiral superfields have
been taken into account, supplemented with the utility of the
equations of motion derived from the Lagrangian density (2.2).}
symmetry transformations (2.3). To this end in mind, we first
compute the following super 2-form $$
\begin{array}{lcl}
\tilde F^{(2)} = \tilde d \tilde A^{(1)} \equiv {\displaystyle
\frac{1}{2!}}\; (dZ^M \wedge dZ^N)\; F_{MN}.
\end{array}\eqno(4.3)
$$
The explicit form of the above computation is
$$
\begin{array}{lcl}
&& \tilde d \tilde A^{(1)} = (dx^\mu \wedge dx^\nu) (\partial_\mu {\cal B}_\nu)
- (d \theta \wedge d \theta) (\partial_\theta \bar {\cal F}) - (d \bar\theta
\wedge d \bar\theta) (\partial_{\bar\theta} {\cal F}) \nonumber\\
&& + (dx^\mu \wedge d\theta) [\partial_\mu \bar {\cal F} - \partial_\theta
{\cal B}_\mu]
+ (dx^\mu \wedge d \bar\theta) [\partial_\mu {\cal F} - \partial_{\bar\theta}
{\cal B}_\mu]
 - (d\theta \wedge d \bar\theta) [\partial_\theta {\cal F} +
\partial_{\bar\theta} \bar {\cal F}].
\end{array}\eqno(4.4)
$$ Mathematically, the horizontality restriction on the four (2,
2)-dimensional supermanifold requires the equality (i.e. $\tilde
F^{(2)} = F^{(2)}$) of the above super curvature 2-form with the
ordinary curvature 2-form $F^{(2)} = d A^{(1)} = \frac{1}{2!}
(dx^\mu \wedge dx^\nu)\; F_{\mu\nu}$, defined on the ordinary 2D
Minkowskian manifold. Physically, this requirement implies that
the Abelian U(1) gauge invariant quantity (i.e. electric field $E$
in our case) remains unaffected due to the presence of the
Grassmannian variables of the superfield formulation. In other
words, one has to set equal to zero all the Grassmannian
components of the (anti)symmetric second-rank super tensor
$F_{MN}$ of (4.3). This, imposition, leads to the following
relationship $$
\begin{array}{lcl}
&& R_\mu = \partial_\mu C, \qquad \bar R_\mu = \partial_\mu \bar C,
\qquad S_\mu = \partial_\mu B,\nonumber\\
&& s = \bar s = 0, \qquad B_1 = \bar B_2 = 0, \qquad
\bar B_1 + B_2 = 0,
\end{array}\eqno(4.5)
$$ where we have identified (i.e. $B_2 = B \Rightarrow \bar B_1 =
- B$) the secondary field $B_2$ with the Nakanishi-Lautrup
auxiliary field $B$ of the Lagrangian density (2.2) and we shall
remain consistent with this identification in the whole body of
our present text.

Taking the help of the nilpotent (anti-)BRST symmetry transformations (2.3),
we obtain the expansions for the superfields (4.2) in terms of these
(i.e. $s_{(a)b}$) when we insert the
specific values (4.5) of the secondary fields. This can be mathematically
expressed as
$$
\begin{array}{lcl}
{\cal B}^{(h)}_\mu (x, \theta, \bar\theta) &=&
A_\mu (x) + \theta\; (s_{ab} A_\mu (x)) + \bar\theta \; (s_b A_\mu (x))
 + \theta\;
\bar\theta\; (s_b s_{ab} A_\mu (x)), \nonumber\\
{\cal F}^{(h)} (x, \theta, \bar\theta) &=&
C (x) + \theta\; (s_{ab} C (x)) + \bar\theta \; (s_b C (x)) + \theta\;
\bar\theta\; (s_b s_{ab} C(x)), \nonumber\\
\bar {\cal F}^{(h)} (x, \theta, \bar\theta) &=&
\bar C (x) + \theta\; (s_{ab} \bar C (x)) + \bar\theta \; (s_b \bar C(x))
 + \theta\;
\bar\theta\; (s_b s_{ab} \bar C (x)).
\end{array}\eqno(4.6)
$$ The noteworthy points, at this stage, are\\ (i) after the
application of the horizontality condition, the super 1-form
connection $\tilde A^{(1)}_{(h)} = dZ^M A_M^{(h)} \equiv dx^\mu
{\cal B}_\mu^{(h)} + d \theta \bar {\cal F}^{(h)} + d \bar\theta
{\cal F}^{(h)}$ is such that $\tilde d \tilde A^{(1)}_{(h)} = d
A^{(1)}$,\\ (ii) the above uniform expressions (4.6) for all the
superfields have emerged out because of the fact that we have
taken into the considerations $s_b C = 0$ and  $s_{ab} \bar C =
0$,\\ (iii) the geometrical interpretations for the nilpotent
(anti-)BRST symmetry transformations (and their corresponding
generators), as the translational generators along the
Grassmannian directions of the supermanifold, are evident from
(4.6). Mathematically, they can be expressed as given below (cf.
equation (2.8)) $$
\begin{array}{lcl}
s_{b} \Leftrightarrow Q_b \Leftrightarrow \mbox{Lim}_{\theta \to 0}
{\displaystyle \frac{\partial}{\partial\bar\theta}}, \qquad
s_{ab} \Leftrightarrow Q_{ab} \Leftrightarrow \mbox{Lim}_{\bar \theta \to 0}
{\displaystyle \frac{\partial}{\partial \theta}},
\end{array}\eqno(4.7)
$$ (iv) the nilpotency property of $s_{(a)b}$ (and their
corresponding generators $Q_{(a)b}$) is equivalent to the two
successive translations of any arbitrary superfields along any
particular direction of the supermanifold because
$(\partial/\partial \theta)^2 = 0, (\partial/\partial
\bar\theta)^2 = 0$, and\\ (v) the anticommutativity  $s_b s_{ab} +
s_{ab} s_b = 0$ of $s_{(a)b}$ (and their corresponding generators,
viz. $Q_b Q_{ab} + Q_{ab} Q_b = 0$) is captured by the
corresponding relationship between the translational generators
(i.e. $(\partial/\partial \theta) (\partial/\partial\bar\theta) +
(\partial/\partial \bar\theta) (\partial/\partial \theta) = 0$) on
the supermanifold.\\

\noindent
{\bf 4.2 Super Co-Exterior Derivative and (Anti-)co-BRST Symmetries}\\

\noindent First of all, let us recall our observations (after
equation (2.4)) where we have discussed the decisive features of
the (anti-)co-BRST symmetry transformations. It is obvious that
the nilpotent (anti-)co-BRST symmetry transformations owe their
origin to the nilpotent co-exterior derivative $\delta = - * d *$.
This claim becomes very much transparent in the framework of
superfield approach to BRST formalism. In fact, we shall exploit
here the mathematical power of the super co-exterior derivative
$\tilde \delta = - \star \tilde d \star$ (with $\tilde \delta^2 =
0$) to show the existence of the local, continuous, off-shell
nilpotent and anticommuting (anti-)co-BRST symmetry
transformations \footnote{It will be noted that these symmetry
transformations have been shown [30-35] to be non-local and
non-covariant for the 4D (non-)Abelian interacting gauge theories
with Dirac fields.} for the free 2D 1-form Abelian gauge theory.
Here $\tilde d$ is the super exterior derivative of (4.1) and the
$\star$ corresponds to the Hodge duality operation on the four (2,
2)-dimensional supermanifold (see, e.g. [45] for details).

Let us impose the following dual-horizontality condition
\footnote{We christen this condition as the dual-horizontality
condition (DHC) because the (super) dual(co)-exterior derivatives
(i.e. $\tilde \delta, \delta$) are being exploited in the
restriction (4.8) on the supermanifold.} (DHC) on the four (2,
2)-dimensional supermanifold over which our 2D 1-form free Abelian
gauge theory is considered: $$
\begin{array}{lcl}
\tilde \delta \tilde A^{(1)} = \delta A^{(1)}, \qquad
\tilde \delta = - \star \tilde d \star, \qquad \delta = - * d *,
\end{array}\eqno(4.8)
$$ where $\delta A^{(1)} = (\partial \cdot A)$ on the ordinary 2D
Minkowskian spacetime manifold. This condition has a logical
backing because the gauge-fixing term $(\partial \cdot A)$ is an
on-shell (i.e. $\Box C = 0, \Box \bar C = 0$) gauge (i.e.
(anti-)BRST) invariant quantity on the supermanifold. This is
evident from the fact that $s_b (\partial \cdot A) = \Box C = 0,
s_{ab} (\partial \cdot A) = \Box \bar C = 0$. The l.h.s. of the
above equation can be computed, step-by-step, due to the following
inputs (see, e.g. [45] for details) $$
\begin{array}{lcl}
\star\; (dx^\mu) &=& \varepsilon^{\mu\nu} (dx_\nu \wedge d \theta \wedge d
\bar\theta), \nonumber\\
\star\; (d \theta) &=& {\displaystyle \frac{1}{2!}}\;\varepsilon^{\mu\nu}
(dx_\mu \wedge dx_\nu \wedge  d
\bar\theta), \nonumber\\
\star\; (d \bar\theta) &=& {\displaystyle \frac{1}{2!}}\;\varepsilon^{\mu\nu}
(dx_\mu \wedge dx_\nu \wedge d \theta),
\end{array}\eqno(4.9)
$$ that are required for the computation of $\star \tilde A^{(1)}$
which is present in $- \star \tilde d \star \tilde A^{(1)}$.
Mathematically, this step, corresponding to the resulting super
3-form, can be expressed  as $$
\begin{array}{lcl}
\star\; \tilde A^{(1)} &=& \varepsilon^{\mu\nu}\;
(dx_\nu \wedge d\theta \wedge
d \bar\theta)\; {\cal B}_\mu
+ {\displaystyle \frac{1}{2!}}\; \varepsilon^{\mu\nu}\;
(dx_\mu \wedge dx_\nu \wedge d\bar\theta)\; \bar {\cal F} \nonumber\\
&+& {\displaystyle \frac{1}{2!}}\; \varepsilon^{\mu\nu}\;
(dx_\mu \wedge dx_\nu \wedge d\theta)\; {\cal F}.
 \end{array}\eqno(4.10)
$$ The application of the super exterior derivative $\tilde d$ on
the above super 3-form makes it a super 4-form. The latter is the
maximum degree of the super-form that could be supported by the
four (2, 2)-dimensional supermanifold. The explicit expression for
it is \footnote{It will be noted that all the super forms having
three wedge products of the spacetime differentials (e. g. $dx_\mu
\wedge dx_\nu \wedge dx_\lambda$) as well as the Grassmannian
differentials (e.g. $d\theta \wedge d\theta \wedge d\theta)$) have
been dropped in the above computation and this will be followed in
the full body of our present text.} $$
\begin{array}{lcl}
&&\tilde d\; \star\; \tilde A^{(1)} = \varepsilon^{\mu\nu}\;
(dx_\rho \wedge dx_\nu \wedge d\theta \wedge d\bar\theta)\;
(\partial^\rho {\cal B}_\mu) \nonumber\\ && - {\displaystyle
\frac{1}{2!}\; \varepsilon^{\mu\nu} (dx_\mu \wedge dx_\nu \wedge
d\theta \wedge d\bar\theta)\; (\partial_\theta \bar {\cal F}) -
\frac{1}{2!}\; \varepsilon^{\mu\nu} (dx_\mu \wedge dx_\nu \wedge
d\theta \wedge d \theta)\; (\partial_\theta {\cal F})}\nonumber\\
&& {\displaystyle - \frac{1}{2!}\; \varepsilon^{\mu\nu} (dx_\mu
\wedge dx_\nu \wedge d\bar\theta \wedge d\bar\theta)\;
(\partial_{\bar\theta} \bar {\cal F}) - \frac{1}{2!}\;
\varepsilon^{\mu\nu} (dx_\mu \wedge dx_\nu \wedge d\theta \wedge
d\bar\theta)\; (\partial_{\bar\theta} {\cal F})}.
\end{array}\eqno(4.11)
$$ The application of a single $(- \star)$ on it would generate a
super zero-form on the four (2, 2)-dimensional supermanifold as
illustrated below (see, e.g. [45] for details)$$
\begin{array}{lcl}
\tilde \delta \tilde A^{(1)} \equiv - \star \tilde d \star \tilde
A^{(1)} = (\partial \cdot {\cal B}) - \partial_\theta \bar {\cal
F} - \partial_{\bar\theta} {\cal F} - s^{\theta\theta}
(\partial_\theta {\cal F}) - s^{\bar\theta\bar\theta}
(\partial_{\bar\theta} \bar {\cal F}),
\end{array}\eqno(4.12)
$$ where we have used, besides $\varepsilon^{\mu\nu}
\varepsilon_{\mu\nu} = - 2!, \varepsilon^{\mu\rho}
\varepsilon_{\nu\rho} = - \delta^\mu_\nu$ etc., the following
duality relations on the supermanifold (see. e.g. [45] for
details) $$
\begin{array}{lcl}
\star\; (dx_\mu \wedge dx_\nu \wedge d\theta \wedge d\bar\theta)
&=& \varepsilon_{\mu\nu}, \nonumber\\ \star\; (dx_\mu \wedge
dx_\nu \wedge d\theta \wedge d\theta) &=& \varepsilon_{\mu\nu}\;
s^{\theta\theta}, \nonumber\\ \star\; (dx_\mu \wedge dx_\nu \wedge
d\bar \theta \wedge d\bar\theta) &=& \varepsilon_{\mu\nu}\;
s^{\bar\theta\bar\theta}.
\end{array}\eqno(4.13)
$$ Here $s^{\theta\theta}$ and $s^{\bar\theta\bar\theta}$ are the
symmetric spacetime independent parameters that are required for
the derivation of the exact expression for the double Hodge
duality operations [45].

All the above computations, combined together, lead to the final
equality as given below
$$
\begin{array}{lcl}
(\partial \cdot {\cal B})
- \partial_\theta \bar {\cal F} - \partial_{\bar\theta} {\cal F}
- s^{\theta\theta} (\partial_\theta {\cal F}) - s^{\bar\theta\bar\theta}
(\partial_{\bar\theta} \bar {\cal F}) = (\partial \cdot A).
\end{array} \eqno(4.14)
$$ It is clear that the coefficients of the Grassmannian dependent
symmetric parameters $s^{\theta\theta}$ and
$s^{\bar\theta\bar\theta}$ would be set equal to zero because
there are no such terms on the r.h.s. of (4.14). This entails upon
the following conditions on the secondary fields, namely; $$
\begin{array}{lcl}
\partial_\theta {\cal F} = 0 \Rightarrow \bar B_1 = 0, \quad \bar s = 0,
\qquad \partial_{\bar\theta} \bar {\cal F} = 0 \Rightarrow B_2 =
0, \quad  s = 0.
\end{array} \eqno(4.15)
$$ The above substitutions in the expansion for the fermionic
superfields ${\cal F}$ and $\bar{\cal F}$ lead to the following
expressions for the reduced form of the above fields $$
\begin{array}{lcl}
{\cal F} (x, \theta, \bar\theta) \;\;\to \;\;{\cal F}^{(r)} (x,
\bar\theta) &=& C (x) + i\; \bar\theta\; B_1, \nonumber\\ \bar
{\cal F} (x, \theta, \bar\theta)\;\; \to \;\; \bar {\cal F}^{(r)}
(x, \theta) &=& \bar C (x) + i\; \theta \;\bar B_2.
\end{array} \eqno(4.16)
$$ Ultimately, the final equality, that incorporates the above
information, is $$
\begin{array}{lcl}
(\partial \cdot {\cal B}) - \partial_\theta \bar {\cal F}^{(r)}
- \partial_{\bar\theta} {\cal F}^{(r)} = (\partial \cdot A).
\end{array} \eqno(4.17)
$$ This equation leads to the following restrictions on the
component fields of the expansions in (4.2) when the explicit
expressions for (4.2) and (4.16) are substituted in it, namely; $$
\begin{array}{lcl}
B_1 + \bar B_2 = 0, \qquad (\partial \cdot \bar R) = 0, \qquad
(\partial \cdot R) = 0, \qquad (\partial \cdot S) = 0.
\end{array} \eqno(4.18)
$$
Making the following judicious choices
$$
\begin{array}{lcl}
R_\mu = - \varepsilon_{\mu\nu} \partial^\nu \bar C, \quad
\bar R_\mu = - \varepsilon_{\mu\nu} \partial^\nu C, \quad
S_\mu = + \varepsilon_{\mu\nu} \partial^\nu {\cal B}, \quad
B_1 = - {\cal B}, \quad \bar B_2 = {\cal B},
\end{array} \eqno(4.19)
$$ it can be seen that all the superfields of (4.2) can be
expressed, in terms of the nilpotent (anti-)co-BRST symmetry
transformations (2.4), as $$
\begin{array}{lcl}
{\cal B}^{(dh)}_\mu (x, \theta, \bar\theta) &=&
A_\mu (x) + \theta\; (s_{ad} A_\mu (x)) + \bar\theta \; (s_d A_\mu (x))
 + \theta\;
\bar\theta\; (s_d s_{ad} A_\mu (x)), \nonumber\\
{\cal F}^{(dh)} (x, \theta, \bar\theta) &=&
C (x) + \theta\; (s_{ad} C (x)) + \bar\theta \; (s_d C (x)) + \theta\;
\bar\theta\; (s_d s_{ad} C(x)), \nonumber\\
\bar {\cal F}^{(dh)} (x, \theta, \bar\theta) &=&
\bar C (x) + \theta\; (s_{ad} \bar C (x)) + \bar\theta \; (s_d \bar C(x))
 + \theta\;
\bar\theta\; (s_d s_{ad} \bar C (x)).
\end{array}\eqno(4.20)
$$ In the above expansion, we have taken into account $s_d \bar C
= 0, s_{ad} C  = 0$. Furthermore, a close look at the above
expansion provides the geometrical interpretation for the
nilpotent symmetry transformations $s_{(a)d}$ (and their
corresponding nilpotent generators) as the translational
generators along the Grassmannian directions of the four (2,
2)-dimensional supermanifold in exactly the same fashion (cf.
(4.7)) as that for the nilpotent (anti-)BRST symmetry
transformations (and their corresponding nilpotent generators).
However, there are clear-cut differences between the HC and DHC
which generate them. The key role is played, in this connection,
by the nature of the expansions of the fermionic superfields
${\cal F}$ and $\bar {\cal F}$ (cf. (4.6) and (4.20)). Whereas,
after the application of HC, the superfields ${\cal F}$ and $\bar
{\cal F}$ become chiral and anti-chiral, respectively, the same
superfields convert themselves to anti-chiral and chiral
superfields after the application of the DHC. Furthermore, the
results obtained due to the application of HC are mathematically
{\it exact} but this is not the situation with the DHC. In the
case of the latter, one has to make a {\it judicious} choice (cf.
(4.19)) for the solutions to the restrictions that emerge (cf.
(4.18)).\\

\noindent
{\bf 4.3 Super Laplacian Operator and Equations of Motion}\\

\noindent As remarked earlier in Sec. 2, the consequence of the
application of the ordinary Laplacian operator $\Delta$ on the
1-form gauge field (i.e. $\Delta A^{(1)} = dx^\mu \Box A_\mu = 0$)
is the equation of motion $\Box A_\mu = 0$ that emerges from the
gauge-fixed Lagrangian density (2.2). One would expect, therefore,
that the action of the super Laplacian operator on the super
1-form connection $\tilde A^{(1)}$ would lead to the derivation of
the equations of motion for all the basic fields $A_\mu, C, \bar
C$ as well as the auxiliary fields $B, {\cal B}$. With the
theoretical arsenal of the definition of the Hodge duality $\star$
operation on the four (2, 2)-dimensional supermanifold [45], we
demonstrate, in this Subsec., that all the other equations of
motion $\Box C = 0, \Box \bar C = 0, \Box B = 0, \Box {\cal B} =
0$ (and their off-shoots $\Box E = 0, \Box (\partial \cdot A) =
0$) emerge from a single restriction on the gauge superfield of
the above supermanifold which owes its origin to the (super)
Laplacian operators that are defined on the (super) spacetime
manifolds.

To corroborate the above assertion, we begin with the following condition
on the four (2, 2)-dimensional supermanifold
$$
\begin{array}{lcl}
\tilde \Delta \tilde A^{(1)} = \Delta A^{(1)} = 0, \qquad \Delta
A^{(1)} = (d \delta + \delta d) A^{(1)} \equiv dx^\mu \Box A_\mu =
0,
\end{array} \eqno(4.21)
$$ where the super Laplacian operator $\tilde \Delta = (\tilde d
\tilde \delta + \tilde \delta \tilde d)$ and super (co-)exterior
derivatives $(\tilde \delta)\tilde d$ (with $\tilde d^2 = 0,
\tilde \delta^2 = 0$) are defined earlier. It is clear from the
r.h.s. of (4.21) that (i) we obtain the equation of motion for the
gauge field (i.e. $\Box A_\mu = 0$) due to the ordinary Laplacian
operator when we demand that the Laplace equation $\Delta A^{(1)}
= 0$ should be satisfied, and (ii) the restriction $\Delta A^{(1)}
= 0 \Rightarrow \Box A_\mu = 0$ is an on-shell ($\Box C = 0, \Box
\bar C = 0$) gauge (i.e. (anti-)BRST) invariant quantity because
$s_{b} (\Box A_\mu) = \partial_\mu (\Box C) = 0, s_{ab} (\Box
A_\mu) =
\partial_\mu (\Box \bar C) = 0$. Therefore, its invariance on the
supermanifold is physically correct. In other words, we require
that the equation of motion $\Box A_\mu = 0$ is unaffected due to
the presence of the Grassmannian variables in the superfield
formulation of the theory.

To compute the exact expression for the l.h.s. (i.e. $\tilde d
\tilde \delta \tilde A^{(1)} + \tilde \delta \tilde d \tilde
A^{(1)}$) of (4.21), let us take the help of our earlier
computation in (4.4) (i.e. $\tilde d \tilde A^{(1)}$) and (4.12)
(i.e. $\tilde \delta \tilde A^{(1)}$). The simpler computation of
$\tilde d \tilde \delta \tilde A^{(1)}$, in its full blaze of
glory, is $$
\begin{array}{lcl}
\tilde d \tilde \delta \tilde A^{(1)} &=& (dx^\mu) \Bigl
[\;\partial_\mu (\partial \cdot {\cal B}) - (\partial_\mu
\partial_\theta \bar {\cal F} + \partial_\mu \partial_{\bar\theta}
{\cal F}) - s^{\theta\theta} (\partial_\mu \partial_\theta {\cal
F}) - s^{\bar\theta\bar\theta} (\partial_\mu \partial_{\bar\theta}
\bar {\cal F})\;\Bigr ] \nonumber\\ &+& (d \theta)\; \bigl
[\;\partial_\theta (\partial \cdot {\cal B}) - \partial_\theta
\partial_{\bar\theta} {\cal F} - s^{\bar\theta\bar\theta}\;
\partial_\theta\partial_{\bar\theta} \bar {\cal F} \;\bigr ]
\nonumber\\ &+& (d \bar\theta)\; \bigl [\;\partial_{\bar\theta}
(\partial \cdot {\cal B}) - \partial_{\bar\theta}
\partial_{\theta} \bar {\cal F} - s^{\theta\theta}\;
\partial_{\bar\theta}\partial_{\theta} {\cal F} \;\bigr ].
\end{array} \eqno(4.22)
$$ In the above computation, the simple definition of $\tilde d$
(cf (4.1)) has been used along with the nilpotency properties of
the Grassmannian partial derivatives (viz. $(\partial_\theta)^2  =
0$ and $(\partial_{\bar\theta})^2 = 0$). Let us now compute,
step-by-step, the more complicated term $\tilde \delta \tilde d
\tilde A^{(1)} = - \star \tilde d \star (\tilde d \tilde A^{(1)})$
where we take the help of (4.4) and $\tilde \delta = - \star
\tilde d \star$. In the first step, we compute $\star (\tilde d
\tilde A^{(1)})$ as $$
\begin{array}{lcl}
\star (\tilde d \tilde A^{(1)}) &=&
\varepsilon^{\mu\rho}\; (d\theta \wedge
d \bar\theta)\; (\partial_\mu {\cal B}_\rho)
\;+ \;\varepsilon^{\mu\rho}\; (d x_\rho \wedge
d \theta)\; (\partial_\mu {\cal F} - \partial_{\bar\theta} {\cal B}_\mu)
\nonumber\\
&+& \varepsilon^{\mu\rho} (d x_\rho \wedge
d \bar\theta)\; (\partial_\mu \bar {\cal F} - \partial_{\theta} {\cal B}_\mu)
- {\displaystyle \frac{1}{2!}}\; \varepsilon^{\mu\nu} (dx_\mu \wedge dx_\nu)
(\partial_\theta {\cal F} + \partial_{\bar\theta} \bar {\cal F})
\nonumber\\
&-& {\displaystyle \frac{1}{2!}}\; s^{\theta\theta}\;
\varepsilon^{\mu\nu} (dx_\mu \wedge dx_\nu)
\;(\partial_\theta \bar {\cal F})
- {\displaystyle \frac{1}{2!}}\; s^{\bar\theta\bar\theta}\;
\varepsilon^{\mu\nu} (dx_\mu \wedge dx_\nu)
\;(\partial_{\bar\theta} {\cal F}).
\end{array} \eqno(4.23)
$$ In the derivation of the above expression, the $\star$
operations that have been used, are [45] $$
\begin{array}{lcl}
&&\star\; (dx^\mu \wedge dx^\rho) = \varepsilon^{\mu\rho}\;
(d\theta \wedge d\bar\theta), \;\;\;\qquad \;\;\;\star\; (dx^\mu
\wedge d\theta) = \varepsilon^{\mu\rho} (dx_\rho \wedge
d\bar\theta), \nonumber\\ && \star\; (dx^\mu \wedge d\bar\theta) =
\varepsilon^{\mu\rho} (dx_\rho \wedge d\theta), \;\;\;\;\qquad
\;\;\;\star\; (d\theta \wedge d\bar\theta) = \frac{1}{2!}\;
\varepsilon^{\mu\nu} (dx_\mu \wedge dx_\nu), \nonumber\\
&&\star\;(d\theta \wedge d\theta) = \frac{1}{2!}\;
s^{\theta\theta}\; \varepsilon^{\mu\nu}\; (dx_\mu \wedge dx_\nu),
\quad \star\;(d\bar\theta \wedge d\bar\theta) = \frac{1}{2!}\;
s^{\bar\theta\bar\theta}\; \varepsilon^{\mu\nu}\; (dx_\mu \wedge
dx_\nu).
\end{array}\eqno(4.24)
$$ We have taken the above expressions for the $\star$ operation
on the super 2-forms from [45] just for the sake of this paper to
be self-contained.

The application of a $\tilde d = dx^\sigma \partial_\sigma + d
\theta \partial_\theta + d \bar\theta \partial_{\bar\theta}$ on
the above expression makes it a super 3-form. In this computation,
all the super forms with the wedge products like $(dx_\mu \wedge
dx_\nu \wedge dx_\lambda)$, $(d\theta \wedge d\theta \wedge
d\theta)$, etc., are to be dropped because the four (2,
2)-dimensional supermanifold cannot support such kind of forms.
Physically relevant super 3-form, that emerges from the above
operation, is $$
\begin{array}{lcl}
\tilde d\;\star\;\tilde d \; \tilde A^{(1)} = L + M + N,
\end{array}\eqno(4.25)
$$
where the exact expressions for L, M and N are
$$
\begin{array}{lcl}
L &=& \varepsilon^{\mu\rho}\; (dx_\sigma \wedge d\theta \wedge d\bar\theta)
\;(\partial^\sigma \partial_\mu {\cal B}_\rho)
+ \varepsilon^{\mu\rho}\; (dx_\sigma \wedge d x_\rho \wedge d\bar\theta)\;
(\partial^\sigma\; [\;
\partial_\mu \bar {\cal F} - \partial_\theta {\cal B}_\mu\;])
\nonumber\\
&+& \varepsilon^{\mu\rho}\; (dx_\sigma \wedge dx_\rho \wedge d\theta)\;
(\partial^\sigma\; [\;
\partial_\mu {\cal F} - \partial_{\bar\theta} {\cal B}_\mu\;])
\end{array}\eqno(4.26)
$$
$$
\begin{array}{lcl}
M &=& \varepsilon^{\mu\rho}\; (dx_\rho \wedge d\theta \wedge d
\bar\theta)\; (\partial_{\theta} \partial_\mu \bar {\cal F}) +
\varepsilon^{\mu\rho}\; (dx_\rho \wedge d\theta \wedge d \theta)\;
(\partial_{\theta} \partial_\mu  {\cal F} -\partial_{\theta}
\partial_{\bar\theta} {\cal B}_\mu)\nonumber\\
&-& {\displaystyle \frac{1}{2!}}\; \varepsilon^{\mu\nu}
(dx_\mu \wedge dx_\nu \wedge d \theta)\;
\bigl [\;\partial_{\theta} \partial_{\bar \theta}\bar {\cal F}
+ s^{\bar\theta\bar\theta}
\partial_{\theta} \partial_{\bar\theta} {\cal F} \;\bigr ].
\end{array}\eqno(4.27)
$$
$$
\begin{array}{lcl}
N &=&
\varepsilon^{\mu\rho}\; (dx_\rho \wedge d\theta \wedge d \bar\theta)\;
(\partial_{\bar\theta} \partial_\mu {\cal F})
+ \varepsilon^{\mu\rho}\; (dx_\rho \wedge d\bar\theta \wedge d \bar\theta)\;
(\partial_{\bar\theta} \partial_\mu \bar {\cal F} -\partial_{\bar\theta}
\partial_\theta {\cal B}_\mu)\nonumber\\
&-& {\displaystyle \frac{1}{2!}}\; \varepsilon^{\mu\nu} \;
(dx_\mu \wedge dx_\nu \wedge d \bar\theta)\;
\bigl [\;\partial_{\bar\theta} \partial_\theta {\cal F} + s^{\theta\theta}
\partial_{\bar\theta} \partial_\theta \bar {\cal F} \;\bigr ].
\end{array}\eqno(4.28)
$$ We are now all set to apply a single ($- \star$) operation on
the above equations which will yield a super 1-form $\tilde \delta
\tilde d \tilde A^{(1)} = - \star \tilde d \star (\tilde d \tilde
A^{(1)})$. This final expression is given as follows $$
\begin{array}{lcl}
&&-\; \varepsilon^{\mu\rho} \varepsilon_{\sigma\lambda}\;
(dx^\lambda)\; (\partial^\sigma \partial_\mu {\cal B}_\rho)
\nonumber\\ &&-\; (dx^\mu) \Bigl [\partial_{\bar\theta}
\partial_\mu {\cal F} + \partial_\theta
\partial_\mu \bar {\cal F}
+ s^{\theta\theta}
\bigl (\partial_\theta \partial_\mu {\cal F} - \partial_\theta
\partial_{\bar\theta} {\cal B}_\mu \bigr )
+ s^{\bar\theta\bar\theta}
\bigl (\partial_{\bar\theta} \partial_\mu \bar {\cal F} -
\partial_{\bar\theta}
\partial_{\theta} {\cal B}_\mu \bigr ) \Bigr ]\nonumber\\
&& + \;(d \theta)\; \bigl [ \;\Box \bar {\cal F}  - \partial_\theta
(\partial \cdot {\cal B}) - \partial_{\bar\theta} \partial_\theta
 {\cal F} - s^{\theta\theta} \partial_{\bar\theta} \partial_\theta
\bar {\cal F}\; \bigr ] \nonumber\\
&& +\; (d \bar\theta)\; \bigl [\; \Box  {\cal F}  - \partial_{\bar\theta}
(\partial \cdot {\cal B}) - \partial_{\theta} \partial_{\bar\theta}
\bar {\cal F} - s^{\bar\theta\bar\theta}
\partial_{\theta} \partial_{\bar\theta}
{\cal F}\; \bigr ].
\end{array}\eqno(4.29)
$$ In the above derivation, the key inputs from [45] have been
used for the $\star$ operations on the super 3-forms. These
relevant expressions are $$
\begin{array}{lcl}
&&\star\;(dx_\sigma \wedge d \theta \wedge d\bar\theta) =
\varepsilon_{\sigma\lambda}\; (dx^\lambda), \qquad
\star\;(dx_\sigma \wedge d x_\rho \wedge d\bar\theta) =
\varepsilon_{\sigma\rho}\; (d\theta), \nonumber\\
&&\star\;(dx_\sigma \wedge d x_\rho \wedge d\theta) =
\varepsilon_{\sigma\rho}\; (d\bar\theta), \qquad \star\;(dx_\rho
\wedge d \theta \wedge d\theta) = \varepsilon_{\rho\lambda}\; (d
x^\lambda)\; s^{\theta\theta}, \nonumber\\ &&\star\;(dx_\rho
\wedge d \bar\theta \wedge d\bar\theta) =
\varepsilon_{\rho\lambda}\; (d x^\lambda)\;
s^{\bar\theta\bar\theta}.
\end{array}\eqno(4.30)
$$ For the sake of the completeness of this paper, we have taken
the above expressions from [45]. The exact value of the operation
of the super Laplacian operator $\tilde \Delta$ on the super
1-form $\tilde A^{(1)}$ is the sum of (4.22) and (4.29).

Let us focus on the terms $s^{\theta\theta} (d\theta),
s^{\theta\theta} (d \bar\theta), s^{\bar\theta\bar\theta}
(d\theta)$ and $s^{\bar\theta\bar\theta} (d\bar\theta)$. These are
certainly not present on the r.h.s. of the restriction (4.21). As
a consequence, these should be set equal to zero. Mathematically,
these statements can be expressed as follows $$
\begin{array}{lcl}
- s^{\theta\theta}\; (d \theta)\; [\;\partial_{\bar\theta} \partial_\theta
\bar {\cal F}\;] &=& 0 \;\;\Rightarrow\;\; \bar s = 0, \qquad
- s^{\theta\theta}\; (d \bar \theta)\; [\;\partial_{\bar\theta} \partial_\theta
{\cal F}\;] = 0 \;\;\Rightarrow\;\;  s = 0, \nonumber\\
- s^{\bar\theta\bar\theta}\; (d \theta)\;
[\;\partial_{\theta} \partial_{\bar\theta}
\bar {\cal F}\;] &=& 0 \;\;\Rightarrow\;\; \bar s = 0, \qquad
- s^{\bar\theta\bar\theta}\; (d \bar\theta) \;
[\;\partial_{\theta} \partial_{\bar\theta}
{\cal F}\;] = 0 \;\;\Rightarrow\;\;  s = 0.
\end{array}\eqno(4.31)
$$
The above inputs lead to the reduced form of the fermionic
superfields ${\cal F} \to {\cal F}^{(r)}$ and
$\bar {\cal F} \to \bar {\cal F}^{(r)}$ as
$$
\begin{array}{lcl}
{\cal F}^{(r)} (x, \theta, \bar\theta) &=& C (x) + i\;\theta\;
\bar B_1 (x) + i\;\bar\theta \; B_1 (x) , \nonumber\\ \bar {\cal
F}^{(r)} (x, \theta, \bar\theta) &=& \bar C (x) + i\;\theta\; \bar
B_2 (x) + i\;\bar\theta \; B_2 (x).
\end{array}\eqno(4.32)
$$ It should be noted that the component fields $s(x)$ and $\bar
s(x)$ are no longer present in the above reduced form of the
fermionic superfields. We collect the coefficient of $d\theta$ and
$d \bar\theta$ from (4.22) and (4.29) and set them equal to zero
as given below $$
\begin{array}{lcl}
(d\theta)\;\bigl [ \Box \bar {\cal F}^{(r)}
- \partial_\theta (\partial \cdot {\cal B}) - \partial_{\bar\theta}
\partial_\theta  {\cal F}^{(r)} + \partial_{\theta} (\partial \cdot
{\cal B}) - \partial_\theta \partial_{\bar\theta} {\cal F}^{(r)} \bigr ] &=& 0,
\nonumber\\
(d\bar\theta)\;\bigl [ \Box  {\cal F}^{(r)}
- \partial_{\bar\theta} (\partial \cdot {\cal B}) - \partial_{\theta}
\partial_{\bar\theta} \bar {\cal F}^{(r)}
+ \partial_{\bar\theta} (\partial \cdot
{\cal B}) - \partial_{\bar\theta} \partial_{\theta}
\bar{\cal F}^{(r)} \bigr ] &=& 0.
\end{array}\eqno(4.33)
$$ It is clear that, finally, we obtain the relations $\Box {\cal
F}^{(r)} = 0$ and $\Box \bar {\cal F}^{(r)} = 0$ due to the fact
that $\partial_\theta \partial_{\bar\theta} +
\partial_{\bar\theta} \partial_\theta = 0$. The above condition
entails upon the component secondary fields of (4.32) to obey
$\Box B_1 = 0, \Box \bar B_1 = 0, \Box B_2 = 0, \Box \bar B_2 =
0$. Our earlier identifications (i.e. $B_1 = - \bar B_2 \equiv -
{\cal B}$ and $\bar B_1 = - B_2 \equiv - B$) imply, ultimately,
 that $\Box B = 0$ and $ \Box {\cal B} = 0$. Furthermore, these equations
of motion, in turn, lead to the conclusion
 that $\Box (\partial \cdot A) = 0$ and
$\Box E = 0$ because of the fact that $B = - (\partial \cdot A)$
and ${\cal B} = E$ from the Lagrangian density (2.2). All the
above equations of motion are the ones that can be derived from
the Lagrangian density (2.2) due to the Euler-Lagrange equations
of motion.

We are now well equipped to collect the coefficients of the
differentials $dx^\mu, (dx^\mu) s^{\theta\theta}$ and $(dx^\mu)
s^{\bar\theta\bar\theta}$. It is obvious from the equality in
(4.21) that the coefficients of $(dx^\mu) s^{\theta\theta}$ and
$(dx^\mu) s^{\bar\theta\bar\theta}$ would be set equal to zero.
This statement can be mathematically expressed as $$
\begin{array}{lcl}
- (dx^\mu) \; s^{\theta\theta} \bigl [ (\partial_\mu
\partial_{\theta} + \partial_{\theta} \partial_\mu) {\cal F}^{(r)}
- \partial_\theta
\partial_{\bar\theta} {\cal B}_\mu \bigr ] &=& 0, \nonumber\\
- (dx^\mu) \; s^{\bar\theta\bar\theta} \bigl [ (\partial_\mu
\partial_{\bar\theta} + \partial_{\bar\theta} \partial_\mu) \bar
{\cal F}^{(r)} - \partial_{\bar\theta} \partial_{\theta} {\cal
B}_\mu \bigr ] &=& 0.
\end{array}\eqno(4.34)
$$ A close look at the above equations implies that (for $dx^\mu
s^{\theta\theta} \neq 0, dx^\mu s^{\bar\theta\bar\theta} \neq 0$)
$$
\begin{array}{lcl}
(\partial_\mu \partial_\theta + \partial_\theta \partial_\mu) {\cal F}^{(r)}
+
(\partial_\mu \partial_{\bar\theta} + \partial_{\bar\theta} \partial_\mu)
\bar {\cal F}^{(r)} = 0.
\end{array}\eqno(4.35)
$$ We collect finally the coefficient of the spacetime
differential $(dx^\mu)$ and equate it with the r.h.s. of (4.21).
This is explicitly given as follows $$
\begin{array}{lcl}
&&(dx^\mu)\; \bigl [ \partial_\mu (\partial \cdot {\cal B}) -
(\partial_\mu \partial_\theta + \partial_\theta \partial_\mu) \bar
{\cal F}^{(r)}
-
(\partial_\mu \partial_{\bar\theta} + \partial_{\bar\theta}
\partial_\mu) {\cal F}^{(r)} \bigr ] \nonumber\\ && -
\varepsilon^{\mu\rho} \varepsilon_{\sigma\lambda} (dx^\lambda)
(\partial^\sigma \partial_\mu {\cal B}_\rho) = dx^\mu \Box A_\mu =
0.
\end{array}\eqno(4.36)
$$ The expansion of all the terms of the equation (4.36) implies
the following $$
\begin{array}{lcl}
(dx^\mu) \; \bigl [\Box {\cal B}_\mu - 2 i \partial_\mu (B_1 +
\bar B_2) \bigr ] = (dx^\mu)\; \Box A_\mu = 0.
\end{array}\eqno(4.37)
$$
The consequences, that emerge from the above equation, are
$$
\begin{array}{lcl}
\Box A_\mu = 0, \qquad \Box R_\mu = 0, \qquad \Box \bar R_\mu = 0,
\qquad \Box S_\mu = 0, \qquad B_1 + \bar B_2 = 0.
\end{array}\eqno(4.38)
$$ A couple of comments are in order now. First, in the above
equation, the last entry is due to the actual restriction $- 2 i
\partial_\mu (B_1 + \bar B_2) = 0$. We have chosen, however, the
solution to the above restriction as $B_1 + \bar B_2 = 0$ because
of our earlier experience with the identification $B_1 = - \bar
B_2 = - {\cal B}$. Finally, in a similar fashion, the restriction
in (4.35) implies that $\partial_\mu (\bar B_1  + B_2) = 0$. Once
again, we have chosen the solution to this restriction as $\bar
B_1 = - B_2 = - B$ due to our earlier experience with such an
identification. With the above inputs, it is clear that the
reduced form of the fermionic superfields in (4.32), are $$
\begin{array}{lcl}
{\cal F}^{(r)} (x, \theta, \bar\theta) &=& C (x) - i\;\theta\;
B(x) + i\;\bar\theta \; {\cal B} (x) , \nonumber\\ \bar {\cal
F}^{(r)} (x, \theta, \bar\theta) &=& \bar C (x) - i\;\theta\;
{\cal B} (x) + i\;\bar\theta \; B (x).
\end{array}\eqno(4.39)
$$ Finally, it is worth pointing out that all the restrictions
$\Box R_\mu = 0, \Box \bar R_\mu = 0$ and $\Box S_\mu = 0$ are
satisfied  due to the validity of equations of motions $\Box C =
0, \Box \bar C = 0, \Box B = 0$ and $\Box {\cal B} = 0$. To make
this point clearer, it can be seen that for the (anti-)BRST
symmetry transformations, we found that $R_\mu = \partial_\mu C,
\bar R_\mu = \partial_\mu \bar C$ and $S_\mu = \partial_\mu B$
(cf. (4.5)). All these values satisfy the restrictions listed in
(4.38). In an exactly similar manner, all the choices that were
made for the derivation of the (anti-)co-BRST symmetry
transformations, namely; $R_\mu = - \varepsilon_{\mu\nu}
\partial^\nu \bar C, \bar R_\mu = - \varepsilon_{\mu\nu}
\partial^\nu  C, S_\mu = + \varepsilon_{\mu\nu} \partial^\nu {\cal
B}$ (cf. (4.19)) do satisfy the restrictions listed in (4.38). In
a more sophisticated mathematical language, the Hodge decomposed
version of the 2D vector secondary (fermionic/bosonic) fields,
chosen in a specific way, namely; $$
\begin{array}{lcl}
R_\mu = \partial_\mu C - \varepsilon_{\mu\nu} \partial^\nu \bar C,
\quad \bar R_\mu = \partial_\mu \bar C - \varepsilon_{\mu\nu}
\partial^\nu C, \quad S_\mu = \partial_\mu B + \varepsilon_{\mu\nu}
\partial^\nu {\cal B},
\end{array}\eqno(4.40)
$$ do satisfy the restrictions listed in (4.38) due to the
equations of motion $\Box C = \Box \bar C = 0, \Box B = \Box {\cal
B} = 0$. In some sense, the above decomposition itself shows that
there exist nilpotent (anti-)BRST and (anti-)co-BRST symmetry
transformations for the free 1-form Abelian gauge theory. This is
due to the fact that, modulo relative signs, the decomposition in
(4.40) is a unique decomposition for the vector fields in two (1 +
1)- dimensions.

Before we wrap up this Subsec., we would like to remark, by taking
the help of the discussions after (3.8), that the conditions
$(\partial \cdot A) = 0$ and $\varepsilon^{\mu\nu} \partial_\mu
A_\nu = 0$ provide the solutions to the equations of motion $\Box
A_\mu = 0$ in two (1 + 1)-dimensions of spacetime. These solutions
emerge due to the symmetry considerations of the present theory.
In an exactly similar fashion, one can provide solutions to the
restrictions $\Box R_\mu = 0, \Box \bar R_\mu = 0$ and $\Box S_\mu
= 0$. These solutions can be enumerated as (i) $(\partial \cdot R)
= 0, \varepsilon^{\mu\nu} \partial_\mu R_\nu = 0$, (ii) $(\partial
\cdot \bar R) = 0, \varepsilon^{\mu\nu} \partial_\mu \bar R_\nu =
0$, and (iii) $(\partial \cdot S) = 0, \varepsilon^{\mu\nu}
\partial_\mu S_\nu = 0$, respectively. Furthermore, the above
solutions, with the help of the Hodge decompositions (4.40),
automatically imply that $\Box C = 0, \Box \bar C = 0, \Box B = 0$
and $\Box {\cal B} = 0$. This is another way to provide the
mathematically and physically beautiful solutions to the
restrictions (4.38) in the two (1 + 1)-dimensions of spacetime.\\

\noindent {\bf 4.4 Topological Aspects: Superfield Formulation}\\

\noindent In this Subsec., we shall express the topological
features of the 2D free 1-form Abelian gauge theory in the
language of the superfield approach to BRST formalism. To this
goal in mind, let us focus on the form of the Lagrangian density
expressed as a sum of the terms that are (anti-)BRST exact and
(anti-)BRST co-exact. These forms are illustrated in their various
guises in (3.1) and (3.5). To express these forms in the language
of the superfields, defined on the four (2, 2)-dimensional
supermanifold, we have to take the help of the super expansions of
the superfields expressed by the equations (4.6) and (4.20). In
fact, it can be seen that the Lagrangian density (2.1) can be
expressed as a total derivative of a specific combinations of
composite superfields with respect to the Grassmannian variable
$\theta$. Mathematically, this statement can be succinctly written
as [51] $$
\begin{array}{lcl}
{\displaystyle {\cal L}_b = +\;\frac{i}{2}\;\mbox
{Lim}_{\bar\theta \to 0}\; \frac{\partial}{\partial \theta}\;
\Bigl [\; \Bigl (\varepsilon^{\mu\nu} \partial_\mu {\cal
B}_\nu^{(dh)} \Bigr )\; \bar {\cal F}^{(dh)} + \Bigl (\partial
\cdot {\cal B}^{(h)} \Bigr )\; {\cal F}^{(h)} \Bigr ]},
\end{array}\eqno(4.41)
$$ where, for our purposes, the expansions of the superfields,
after the application of HC and DHC (cf. (4.6) and (4.20)), can be
re-expressed as $$
\begin{array}{lcl}
{\cal B}^{(h)}_\mu (x, \theta, \bar\theta) &=& A_\mu (x) +
\theta\; (\partial_\mu \bar C (x)) + \bar\theta \; (\partial_\mu C
(x)) + i\;\theta\; \bar\theta\; (\partial_\mu B (x)), \nonumber\\
{\cal F}^{(h)} (x, \theta) &=& C (x) - i\; \theta\; B (x), \qquad
\bar {\cal F}^{(h)} (x, \bar\theta) = \bar C (x) + i\;\bar\theta
\;B (x),
\end{array}\eqno(4.42)
$$
$$
\begin{array}{lcl}
{\cal B}^{(dh)}_\mu (x, \theta, \bar\theta) &=& A_\mu (x) -
\theta\; \varepsilon_{\mu\nu} \partial^\nu C (x) -  \bar\theta \;
\varepsilon_{\mu\nu} \partial^\nu \bar C (x)
 + i\;\theta\;
\bar\theta\; \varepsilon_{\mu\nu} \partial^\nu {\cal B} (x),
\nonumber\\ {\cal F}^{(dh)} (x, \bar\theta) &=& C (x) -
i\;\bar\theta \; {\cal B} (x), \qquad \bar {\cal F}^{(dh)} (x,
\theta) = \bar C (x) + i\; \theta\; {\cal B} (x).
\end{array}\eqno(4.43)
$$ For the on-shell nilpotent symmetry transformations, given in
(3.2) and (3.3), the above expansions can be re-written by
inserting the values of the auxiliary fields $B = - (\partial
\cdot A)$ and ${\cal B} = E$ that emerge from the Lagrangian
density (2.2). Precisely speaking, for the derivation of the
on-shell nilpotent symmetry transformations for the gauge and
(anti-)ghost fields, one has to invoke (anti-)chiral superfields
on the three (2, 1)-dimensional sub-manifold(s) of the four (2,
2)-dimensional supermanifold [49-51]. However, for our
discussions, we shall take the appropriate form of the expansions
of the superfields from the above equations. These are
re-expressed as follows $$
\begin{array}{lcl}
{\cal B}^{(h)}_\mu (x, \theta, \bar\theta) &=& A_\mu (x) +
\theta\; (\partial_\mu \bar C (x)) + \bar\theta \; (\partial_\mu C
(x)) - i\;\theta\; \bar\theta\; (\partial_\mu  (\partial \cdot A)
(x)), \nonumber\\ {\cal F}^{(h)} (x, \theta) &=& C (x) + i\;
\theta\; (\partial \cdot A) (x), \qquad \bar {\cal F}^{(h)} (x,
\bar\theta) = \bar C (x) - i\;\bar\theta \; (\partial \cdot A)
(x),
\end{array}\eqno(4.44)
$$ $$
\begin{array}{lcl}
{\cal B}^{(dh)}_\mu (x, \theta, \bar\theta) &=& A_\mu (x) -
\theta\; \varepsilon_{\mu\nu} \partial^\nu C (x) -  \bar\theta \;
\varepsilon_{\mu\nu} \partial^\nu \bar C (x)
 + i\;\theta\;
\bar\theta\; \varepsilon_{\mu\nu} \partial^\nu E (x), \nonumber\\
{\cal F}^{(dh)} (x, \bar\theta) &=& C (x) - i\;\bar\theta \; E
(x), \qquad \bar {\cal F}^{(dh)} (x, \theta) = \bar C (x) + i\;
\theta\; E (x).
\end{array}\eqno(4.45)
$$ The expressions, quoted in (4.44) and (4.45), are to be
utilized in the equation (4.41) for the derivation of the
Lagrangian density. It is elementary to check that, modulo a total
derivative, the Lagrangian density (2.1) emerges from the
Lagrangian density (4.41). To be precise, the expression (4.41) is
the analogue of the Lagarangian densities (3.1) and (3.5) which
are expressed in terms of (i) the nilpotent symmetry operators
$\tilde s_{ab}$ and $\tilde s_{ad}$ (cf. (3.1)), and (ii) the
corresponding nilpotent symmetry generators $\tilde Q_{ab}$ and
$\tilde Q_{ad}$ (cf. (3.5)).

The Lagrangian densities (3.1) and (3.5) for the new TFTs, that
have been expressed as the sum of BRST exact and co-exact terms in
the language of either the symmetry transformation operators
$\tilde s_b$ and $\tilde s_d$ (or the corresponding nilpotent
symmetry generators $\tilde Q_b$ and $\tilde Q_d$), can be written
as the total derivative with respect to the Grassmannian variable
$\bar\theta$. The precise expression for such a form of the
Lagrangian density, in the framework of the superfield approach to
BRST formalism, is $$
\begin{array}{lcl}
{\displaystyle {\cal L}_b = -\;\frac{i}{2}\; \mbox {Lim}_{\theta
\to 0}\; \frac{\partial}{\partial \bar \theta}\; \Bigl [\;
(\varepsilon^{\mu\nu} \partial_\mu {\cal B}_\nu^{(dh)})\; {\cal
F}^{(dh)} + (\partial \cdot {\cal B}^{(h)})\; \bar {\cal F}^{(h)}
\Bigr ]}.
\end{array}\eqno(4.46)
$$ It can be easily checked that the substitution of the
superfield expansions, given in (4.44) and (4.45), into the above
equation leads to the derivation of the Lagrangian density (2.1),
modulo a total derivative. A close look at the expressions (4.41)
and (4.46) provides the geometrical interpretation for the
Lagrangian density (2.1) of the 2D free 1-form Abelian gauge
theory which happens to be a new TFT. Geometrically, this
Lagrangian for the 2D gauge theory, defined on the ordinary 2D
Minkowskian spacetime manifold, corresponds to the translation of
the sum of the composite superfields $ (\partial \cdot {\cal
B}^{(h)}) {\cal F}^{(h)}$ and $(\varepsilon^{\mu\nu}
\partial_\mu {\cal B}^{(dh)}) \bar {\cal F}^{(dh)}$ along the
$\theta$-direction of the four (2, 2)-dimensional supermanifold
over which the above superfields have been defined. In an exactly
analogous manner, the geometrical interpretation for the
Lagrangian density (4.46) can be provided in the language of the
translations of the combination of the superfields $(\partial
\cdot {\cal B}^{(h)}) \bar {\cal F}^{(h)}$ and
$(\varepsilon^{\mu\nu} \partial_\mu {\cal B}_\nu^{(dh)}) {\cal
F}^{(dh)}$ along the $\bar\theta$-direction of the four (2,
2)-dimensional supermanifold. It should be noted here that the
combination of the composite superfields in the Lagrangian
densities (4.41) and (4.46) is such that individually these terms
(i.e. the composite superfields) turn out to be the Lorentz
scalars.

Besides the superfield formulation of the Lagrangian density (2.1)
in terms of translations along $\theta$-direction (cf. (4.41)) and
$\bar\theta$-direction (cf. (4.46)) of the four (2, 2)-dimensional
supermanifold, there is a third way of expressing the starting
Lagrangian density (2.1) in the language of superfield approach to
BRST formulation. This new way of expressing the Lagrangian
density (2.1), modulo a total spactime derivative, is $$
\begin{array}{lcl}
{\cal L}_b = {\displaystyle \frac{i}{4}\; \frac{\partial}{\partial
\bar\theta}\; \frac{\partial}{\partial\theta} \; \Bigl [\;{\cal
B}_\mu^{(h)} {\cal B}^{\mu (h)}\; + \; {\cal B}_\mu^{(dh)} {\cal
B}^{\mu (dh)}\; \Bigr ]}.
\end{array}\eqno(4.47)
$$ A few comments are in order now. First, it can be noted that
the bosonic Lorentz scalar (i.e. ${\cal B}^\mu {\cal B}_\mu$), on
which the Grassmannian derivatives operate, is constructed with
the help of a single bosonic superfield ${\cal B}_\mu$. Second,
the HC and DHC, which are responsible for the derivation of the
nilpotent (anti-)BRST and (anti-)co-BRST symmetry transformations
in the framework of the superfield formulation, play very
important roles in the derivation of the Lagrangian density
(4.47). Third, in the language of the geometry on the four (2,
2)-dimensional supermanifold, the above Lagrangian density is
nothing but a couple of successive translations for the sum of the
composite bosonic Lorentz scalar superfields $({\cal B}_\mu^{(h)}
{\cal B}^{\mu (h)})$ and $({\cal B}_\mu^{(dh)} {\cal B}^{\mu
(dh)})$ along the Grassmannian  $\theta$- and
$\bar\theta$-directions of the four (2, 2)-dimensional
supermanifold, respectively. Finally, the above Lagrangian density
can be expressed in terms of the on-shell nilpotent (anti-)BRST
and (anti-)co-BRST symmetry transformations of (3.2) and (3.3) as
given below $$
\begin{array}{lcl}
{\cal L}_b = {\displaystyle \frac{i}{4}\; \tilde s_b\; \tilde
s_{ab}\; \Bigl [\; (A_\mu (x) A^\mu (x)) \Bigr ] + \frac{i}{4}\;
\tilde s_d\;\tilde s_{ad} \; \Bigl [\; (A_\mu (x) A^\mu (x)) \Bigr
]}.
\end{array}\eqno(4.48)
$$ We would like to emphasize that the above expression for the
starting Lagrangian density is a new one and it has been derived
only due to our understanding of the superfield approach to BRST
formalism. The crucial role in this derivation has been played by
our knowledge of the following beautiful mappings $$
\begin{array}{lcl}
&& \tilde s_{b} A_\mu (x) \Leftrightarrow -\;i\; [ A_\mu (x),
\tilde Q_b ] \Leftrightarrow \mbox{Lim}_{\theta \to 0}
{\displaystyle \frac{\partial}{\partial\bar\theta}} {\cal
B}_\mu^{(h)}, \nonumber\\ && \tilde s_{ab} A_\mu (x)
\Leftrightarrow -\; i\; [ A_\mu (x), \tilde Q_{ab} ]
\Leftrightarrow \mbox{Lim}_{\bar \theta \to 0} {\displaystyle
\frac{\partial}{\partial \theta} {\cal B}_\mu^{(h)}}, \nonumber\\
&& \tilde s_{d} A_\mu (x) \Leftrightarrow -\;i\; [ A_\mu (x),
\tilde Q_d ] \Leftrightarrow \mbox{Lim}_{\theta \to 0}
{\displaystyle \frac{\partial}{\partial\bar\theta}} {\cal
B}_\mu^{(dh)}, \nonumber\\ && \tilde s_{ad} A_\mu (x)
\Leftrightarrow -\; i\; [ A_\mu (x), \tilde Q_{ad} ]
\Leftrightarrow \mbox{Lim}_{\bar \theta \to 0} {\displaystyle
\frac{\partial}{\partial \theta} {\cal B}_\mu^{(dh)}}.
\end{array}\eqno(4.49)
$$ In terms of the conserved and on-shell nilpotent charges
$\tilde Q_r$ (with $r = b, ab, d, ad$), the above Lagrangian
density (4.48) can be re-expressed as $$
\begin{array}{lcl}
{\cal L}_b = - {\displaystyle \;\frac{i}{4}\; \bigl \{ \tilde Q_b,
\; [ \tilde Q_{ab},\; A_\mu (x) A^\mu (x) ] \bigr \} -
\frac{i}{4}\; \bigl \{ \tilde Q_d,\; [ \tilde Q_{ad}, \; A_\mu (x)
A^\mu (x) ]\bigr \}}.
\end{array}\eqno(4.50)
$$ It will be noted that all the expressions for the Lagrangian
density, given in (4.47), (4.48) and (4.50), differ from the
starting Lagrangian density (2.1) by a total spactime derivative
which is equal to : $\frac{1}{2} \;\partial^\mu [\;A_\mu (\partial
\cdot A) + \varepsilon_{\mu\nu} A^\nu E \;]$.

We express the form of the symmetric energy-momentum tensor (3.7)
in the language of the superfield approach to BRST formalism.
However, before accomplishing this goal, let us express the above
symmetric energy momentum tensor in terms of the nilpotent
symmetry transformations given in (3.2) and (3.3). The exact
expression for (3.7) is $$
\begin{array}{lcl}
T_{\mu\nu}^{(s)} = {\displaystyle \frac{i}{2}\; \tilde s_b\;
\tilde s_{ab}\; \Bigl [\; A_\mu  A_\nu  - \frac{1}{2}
\;\eta_{\mu\nu}\; A_\rho A^\rho\; \Bigr ] - \frac{i}{2}\; \tilde
s_d\;\tilde s_{ad} \; \Bigl [\; \varepsilon_{\mu\rho}
\varepsilon_{\nu\sigma} A^\rho A^\sigma + \frac{1}{2}
\;\eta_{\mu\nu}\; A_\rho A^\rho\;
 \Bigr ]}.
\end{array}\eqno(4.51)
$$ Exploiting now the mapping given in (4.49), it is easier to
express the form of the symmetric energy momentum tensor in the
language of the translations of some combination of composite
superfields along the Grassmannian directions of the four (2,
2)-dimensional super manifold. This explicit expression, in terms
of the superfields, is $$
\begin{array}{lcl}
T_{\mu\nu}^{(s)} &=& {\displaystyle \frac{i}{2}\;
\frac{\partial}{\partial \bar\theta}\;
\frac{\partial}{\partial\theta} \; \Bigl [\;{\cal B}_\mu^{(h)}
{\cal B}_{\nu}^{(h)} - \frac{1}{2}\;\eta_{\mu\nu}\; {\cal
B}_\rho^{(h)} {\cal B}^{\rho (h)}} \nonumber\\ &-& {\displaystyle
\varepsilon_{\mu\rho}\; \varepsilon_{\nu\sigma} {\cal B}^{\rho
(dh)}\; {\cal B}^{\sigma (dh)} - \frac{1}{2}\;\eta_{\mu\nu}\;
{\cal B}_\rho^{(dh)} {\cal B}^{\rho (dh)} \Bigr ]},
\end{array}\eqno(4.52)
$$ where the super expansions of the bosonic superfield ${\cal
B}_\mu$ has to be taken from (4.44) and (4.45) which are obtained
after the application of the HC and DHC. Geometrically, the above
equation corresponds to a couple of successive translations of the
combination of the {\it symmetric} composite superfields along the
Grassmannian $\theta$- and $\bar\theta$-directions of the four (2,
2)-dimensional supermanifold, respectively. The above Lorentz
scalar composite superfields are constructed with the help of the
bosonic superfields of the expansions (4.44) and (4.45) {\it
alone} (that have been obtained after the application of HC and
DHC).

The analogue of equations (4.51) and (4.52) can also be written if
we take the help of equations (3.7) and (3.8). Taking into account
the latter, it can be seen that the symmetric energy momentum
tensor can be also expressed as $$
\begin{array}{lcl}
T_{\mu\nu}^{(s)} = \tilde s_b\; (i V_{\mu\nu}^{(1)}) + \tilde s_d
\; (i V_{\mu\nu}^{(2)}) \equiv \tilde s_{ab}\; (i \bar
V_{\mu\nu}^{(1)}) + \tilde s_{ad} \; (i \bar V_{\mu\nu}^{(2)}),
\end{array}\eqno(4.53)
$$ where the explicit expressions for the $V_{\mu\nu}^{(1,2)}$ and
$\bar V_{\mu\nu}^{(1,2)}$, in terms of the local fields of the
Lagrangian density (2.1), are given in (3.8). Exploiting the
mappings given in (4.49), it can be seen that the symmetric
energy-momentum tensor can also be written in the following total
derivative forms $$
\begin{array}{lcl}
T_{\mu\nu}^{(s)} &=& {\displaystyle  \frac{i}{2} \; \mbox
{Lim}_{\theta \to 0}\; \frac{\partial}{\partial\bar\theta} \;
\Bigl [ \; \Bigl \{ \eta_{\mu\nu} \; (\partial \cdot {\cal
B}^{(h)}) - \bigl (
\partial_\mu {\cal B}^{(h)}_\nu + \partial_\nu {\cal B}^{(h)}_\mu \bigr ) \Bigr \}\;
\bar {\cal F}^{(h)} }\nonumber\\ &+& \;\Bigl \{ \eta_{\mu\nu}
\;\varepsilon^{\rho\sigma} \partial_\rho {\cal B}_\sigma^{(dh)}\;
- \bigl (\varepsilon_{\nu\rho}
\partial_\mu {\cal B}^{\rho (dh)} +
\varepsilon_{\mu\rho}
\partial_\nu {\cal B}^{\rho (dh)}
 \bigr ) \Bigr \}\; {\cal F}^{(dh)}\;\Bigr ],
\end{array}\eqno(4.54)
$$
$$
\begin{array}{lcl}
T_{\mu\nu}^{(s)} &=& {\displaystyle - \frac{i}{2} \; \mbox
{Lim}_{\bar\theta \to 0}\; \frac{\partial}{\partial \theta} \;
\Bigl [ \; \Bigl \{ \eta_{\mu\nu} \; (\partial \cdot {\cal
B}^{(h)}) - \bigl (
\partial_\mu {\cal B}^{(h)}_\nu
+ \partial_\nu {\cal B}^{(h)}_\mu \bigr ) \Bigr \}\;
 {\cal F}^{(h)} }\nonumber\\ &+& \;\Bigl \{ \eta_{\mu\nu}
\;\varepsilon^{\rho\sigma} \partial_\rho {\cal B}_\sigma^{(dh)}\;
- \bigl (\varepsilon_{\nu\rho}
\partial_\mu {\cal B}^{\rho (dh)} +
\varepsilon_{\mu\rho}
\partial_\nu {\cal B}^{\rho (dh)}
 \bigr ) \Bigr \}\; \bar {\cal F}^{(dh)}\;\Bigr ],
\end{array}\eqno(4.55)
$$ where the expansions for the superfields have to be inserted
from (4.44) and (4.45) which have been obtained after the
applications of HC and DHC. In the language of the geometry on the
four (2, 2)-dimensional supermanifold, it can be seen that the
symmetrical energy momentum tensor is equivalent to the
translations (i) along the Grassmannian $\bar\theta$-direction of
the supermanifold when a specific combination of the Lorentz
scalar composite superfields (cf. (4.54)) is taken into account,
and (ii) along the Grassmannian $\theta$-direction of the four (2,
2)-dimensional supermanifold when the same combination of the
Lorentz scalar composite superfields (cf. (4.55)) is taken into
account, modulo a relative sign factor.

It is very interesting to point out the fact that the local field
operators $V_{\mu\nu}^{(1, 2)}$ and $\bar V_{\mu\nu}^{(1, 2)}$ can
also be written as the total derivatives of the appropriate
composite superfields (taken from the expansions (4.44) and
(4.45)) with respect to the Grassmannian variables $\theta$ and
$\bar\theta$. The explicit expressions for these operators are $$
\begin{array}{lcl}
V_{\mu\nu}^{(1)} &=& {\displaystyle + \frac{1}{2} \; \mbox
{Lim}_{\bar\theta \to 0}\; \frac{\partial}{\partial \theta} \;
\Bigl [\; {\cal B}_\mu^{(h)}\; {\cal B}_\nu^{(h)} - i\;
\eta_{\mu\nu} \;{\cal F}^{(h)}\;\bar {\cal F}^{(h)}\; \Bigr ]
}\nonumber\\ V_{\mu\nu}^{(2)} &=& - {\displaystyle \frac{1}{2}\;
\mbox {Lim}_{\bar\theta \to 0}\; \frac{\partial}{\partial \theta}
\; \Bigl [\; \varepsilon_{\mu\rho}\;\varepsilon_{\nu\sigma} \;
{\cal B}^{\rho (dh)} \; {\cal B}^{\sigma (dh)} + i\;\eta_{\mu\nu}
\; {\cal F}^{(dh)}\; \bar {\cal F}^{(dh)}\; \Bigr ]},
\end{array}\eqno(4.56)
$$ $$
\begin{array}{lcl}
\bar V_{\mu\nu}^{(1)} &=& - {\displaystyle  \frac{1}{2} \; \mbox
{Lim}_{\theta \to 0}\; \frac{\partial}{\partial\bar\theta} \;
\Bigl [\; {\cal B}_\mu^{(h)}\; {\cal B}_\nu^{(h)} - i\;
\eta_{\mu\nu} \;{\cal F}^{(h)}\;\bar {\cal F}^{(h)}}\; \Bigr
],\nonumber\\ \bar V_{\mu\nu}^{(2)} &=& + {\displaystyle
\frac{1}{2} \; \mbox {Lim}_{\theta \to 0}\;
\frac{\partial}{\partial\bar\theta} \; \Bigl [\;
\varepsilon_{\mu\rho}\;\varepsilon_{\nu\sigma} \; {\cal B}^{\rho
(dh)} \; {\cal B}^{\sigma (dh)} + i\;\eta_{\mu\nu} \; {\cal
F}^{(dh)}\; \bar {\cal F}^{(dh)}\; \Bigr ]},
\end{array}\eqno(4.57)
$$ The key consequences of the above explicit expressions are
related to yet another way of expressing the symmetric
energy-momentum tensor (3.7) in terms of the superfields and
Grassmannian derivatives. For this purpose, the geometrical
mappings given in (4.49), turn out to be quite handy. Taking the
help of (4.56), (4.57) and (4.49), we obtain the following
expressions for the symmetric energy-momentum tensor (3.7),
namely;$$
\begin{array}{lcl}
T_{\mu\nu}^{(s)} &=& {\displaystyle + \frac{i}{2} \;
\frac{\partial}{\partial \bar\theta}\; \frac{\partial}{\partial
\theta} \; \Bigl [\; {\cal B}_\mu^{(h)}\; {\cal B}_\nu^{(h)} - i\;
\eta_{\mu\nu} \;{\cal F}^{(h)}\;\bar {\cal F}^{(h)}}\nonumber\\ &
-& \varepsilon_{\mu\rho}\;\varepsilon_{\nu\sigma} \; {\cal
B}^{\rho (dh)} \; {\cal B}^{\sigma (dh)} - i\;\eta_{\mu\nu} \;
{\cal F}^{(dh)}\; \bar {\cal F}^{(dh)}\; \Bigr ],
\end{array}\eqno(4.58)
$$ $$
\begin{array}{lcl}
T_{\mu\nu}^{(s)} &=& {\displaystyle - \frac{i}{2} \;
\frac{\partial}{\partial \theta}\; \frac{\partial}{\partial \bar
\theta} \; \Bigl [\; {\cal B}_\mu^{(h)}\; {\cal B}_\nu^{(h)} - i\;
\eta_{\mu\nu} \;{\cal F}^{(h)}\;\bar {\cal F}^{(h)}}\nonumber\\ &
-& \varepsilon_{\mu\rho}\;\varepsilon_{\nu\sigma} \; {\cal
B}^{\rho (dh)} \; {\cal B}^{\sigma (dh)} - i\;\eta_{\mu\nu} \;
{\cal F}^{(dh)}\; \bar {\cal F}^{(dh)}\; \Bigr ].
\end{array}\eqno(4.59)
$$ It will be noted that (i) the above expressions are different
from their counterpart in (4.52), (ii) the expression in (4.59) is
not an independent quantity because it can be derived from (4.58)
by exploiting the trivial relationship $\partial_\theta
\partial_{\bar\theta} + \partial_{\bar\theta}
\partial_\theta = 0$, and (iii) the above expressions can be
re-written in terms of the on-shell nilpotent (anti-)BRST and
(anti-)co-BRST symmetry transformations (3.2) and (3.3) as $$
\begin{array}{lcl}
T_{\mu\nu}^{(s)} &=& {\displaystyle + \frac{i}{2}} \; \tilde s_b
\tilde s_{ab}\; \Bigl [\; A_\mu  A_\nu - i\; \eta_{\mu\nu}\; C
\bar C\; \Bigr ] - {\displaystyle \frac{i}{2}}\;\tilde s_d \tilde
s_{ad} \;\Bigl [\; \varepsilon_{\mu\rho}\;\varepsilon_{\nu\sigma}
 A^{\rho}  A^{\sigma} + i\;\eta_{\mu\nu} \; C  \bar C\; \Bigr ].
\end{array}\eqno(4.60)
$$ It is obvious that the above expression is quite different from
its counterpart in (4.51) which has also been expressed in terms
of the on-shell nilpotent symmetry transformations.

We close this subsection with the remark that {\it mathematically}
the Lagrangian density and symmetric energy-momentum tensor of a
TFT, in the framework of the superfield approach to BRST
formalism, can always be expressed as a total derivative with
respect to (i) the Grassmannian variable $\theta$, or (ii) the
Grassmannian variable $\bar\theta$, and/or (iii) a combination of
the Grassmannian variables $\theta$ and $\bar\theta$ together. In
the language of the geometry on the appropriately chosen
supermanifold, the Lagrangian density as well as symmetric energy
momentum tensor, of a given TFT, always correspond to the
translations(s) of the composite superfields along the
Grassmannian directions of the above supermanifold. The above
statements are true for the TFTs which have the mathematical
structure (i.e the form of the Lagrangian density, energy-momentum
tensor etc.) like Witten type TFT.\\

 \noindent {\bf 5. Conclusions}\\

\noindent In our present endeavor, we have mainly focused our
attention on the continuous as well as discrete symmetry
transformations of the Lagrangian density of a given 2D free
1-form Abelian gauge theory in the framework of the BRST
formalism. To be specific, we have shown the existence of the {\it
continuous} nilpotent (anti-)BRST symmetry transformations,
nilpotent (anti-)co-BRST symmetry transformations and a
non-nilpotent (bosonic) symmetry transformation for the Lagrangian
density of the above 2D free 1-form Abelian gauge theory. The
above symmetry transformations (and their corresponding
generators) have been, in turn, shown to possess a deep connection
with the de Rham cohomological operators of the differential
geometry. To establish the above connection in its totality, one
requires the existence of a specific set of {\it discrete}
symmetry transformations in the theory. These symmetries have also
been shown to exist for the 2D free 1-form Abelian gauge theory.
In fact, it is the combination and interplay of the discrete and
continuous symmetry transformations that provides the exact
analogue of the relationship that exists between the (co-)exterior
derivatives $(\delta) d$ (i.e. $\delta = \pm * d *$). Precisely
speaking, the discrete symmetry transformations of the theory
correspond to the Hodge duality $*$ operation of the differential
geometry in the celebrated relationship $\delta = \pm \;*\;d \;*$.

We have chosen here the 2D free Abelian 1-form gauge theory as the
prototype example for the Hodge theory because, for this tractable
field theoretical model, one obtains the local, continuous,
covariant and nilpotent (anti-)BRST and (anti-)co-BRST symmetry
transformations which is not the case for the 4D 1-form free as
well as interacting (non-)Abelian gauge theories. In fact, it
turns out that the above (anti-)co-BRST symmetry transformations
for the 4D theories are non-local and non-covariant [30-35]. One
can obtain the covariant version of the above symmetry
transformations by introducing a specific kind of parameter in the
theory but the nilpotency of the symmetry transformations
disappears. The latter property is restored only for a specific
value of the above parameter [35]. As a consequence, the conserved
charges (which turn out to be the generators for the above
(anti-)co-BRST symmetry transformations) are found to be non-local
for the 4D theories. The above cited problems do not arise for the
2D {\it free} as well as {\it interacting} 1-form Abelian gauge
theory. We have discussed, in our Appendix, these (i.e. nilpotent
(anti-)co-BRST) symmetry transformations associated with the
interacting 2D 1-form Abelian gauge theory where there is a direct
coupling between the U(1) gauge field and the Noether conserved
current constructed with the help of the fermionic Dirac fields.

It is precisely due to the existence of the well-defined on-shell
nilpotent (anti-)BRST and (anti-)co-BRST symmetry transformations
(and their corresponding generators) that the Lagrangian density
of the 2D free 1-form Abelian gauge theory has been able to be
expressed as the sum of two anticommutators that are constructed
with the help of the conserved and on-shell nilpotent (anti-)BRST
and (anti-)co-BRST charges. In other words, the Lagrangian density
turns our to be the sum of the BRST exact and BRST co-exact terms.
This observation, in turn, implies that the Lagrangian density of
the above theory has a form which is similar to the Witten type
topological field theories. Similarly, the symmetric
energy-momentum tensor of the above theory can also be expressed
as the sum of the BRST exact and BRST co-exact terms. However, (i)
the absence of the local topological shift symmetry
transformations, and (ii) the presence of the gauge type (i.e.
BRST and co-BRST) symmetry transformations implies that the
present 2D 1-form Abelian gauge theory is similar to  the Schwarz
type of topological field theories. Thus, we conclude that the
present 2D free 1-form Abelian gauge theory is a {\it new} type of
TFT. Its topological nature has been shown through the application
of the BRST and co-BRST charges on the {\it harmonic} (i.e.
physical) state of the Hodge decomposed state in the quantum
Hilbert space of states. It turns out that both the dynamical
degrees of freedom of the 2D photon ($A_\mu$) field can be gauged
away due to the coexistence of the BRST and co-BRST symmetry
transformations {\it together}. We have established this result
for the case of a single physical photon (harmonic) state that is
created from the physical vacuum of the theory.

One of the key features of the present field theoretical model is
that the model provides a beautiful example for the application of
the techniques involved in the geometrical superfield approach to
BRST formalism. To be specific and precise, the following points,
connected with the superfield formalism, are noteworthy. First,
the nilpotent (anti-)BRST symmetry transformations owe their
origin to the (super) exterior derivatives $(\tilde d) d$ which
are exploited in the horizontality condition defined on the four
(2, 2)-dimensional supermanifold (over which the  2D 1-form
Abelian theory is considered). Second, the nilpotent
(anti-)co-BRST symmetry transformations are found to be deeply
connected with the (super) co-exterior derivatives $(\tilde
\delta) \delta$ which are exploited in the DHC imposed on the
above supermanifold. The (super) Laplacian operators $(\tilde
\Delta) \Delta$, however, lead to the derivation of the equations
of motion for all the fields of the 2D free Abelian gauge theory
from an appropriate restriction (cf. (4.21)) on the above
supermanifold. Thus, we note that all the (super) cohomological
operators play very significant roles in describing various
aspects of the model under consideration. This is why, the 2D free
1-form Abelian gauge theory is a {\it perfect} field theoretical
model for the Hodge theory. Moreover, the topological features of
the model under consideration have also been shown in terms of the
superfields and Grassmannian derivatives.

A very important aspects of our discussion is the significance of
the definition of the Hodge duality $\star$ operation on the four
(2, 2)-dimensional supermanifold [45]. In fact, our present model
provides a very good example where the correctness of the above
Hodge duality $\star$ operation [45] has been tested in a
meaningful manner. It will be noted that the above definition of
the duality plays a key role in the application of the DHC on the
supermanifold. Furthermore, the above definition is required in
the application of the restriction (4.21) on the four (2,
2)-dimensional supermanifold where the super Laplacian operator
$\tilde \Delta$ plays a key role. For the present model, we find
that the definition of the Hodge duality $\star$ operation, on the
appropriately chosen four (2, 2)-dimensional supermanifold, is
correct because it leads to the precise and consistent results
which have been discussed in Subsec. 4.2 and Subsec. 4.3. It is
gratifying to note that, for the model under consideration, the
{\it ordinary} and {\it super} de Rham cohomological operators
blend together in a beautiful manner to lead to some very cute
theoretical results (thereby rendering the present theory to be a
perfect field theoretical model for the Hodge theory).

It would be very interesting endeavor to apply our present ideas
to the physical 4D field theories. In this connection, it is worth
pointing out that, in our earlier works [42-44], we have been able
to show the existence of the local, continuous, covariant and
nilpotent (anti-)co-BRST symmetry transformations for the 4D free
2-form Abelian gauge theory. In fact,  we have been able to prove
that this 4D field theoretical model, in its Lagrangian
formulation, is an example of the Hodge theory [44]. However, we
have not yet been able to establish the role of the super de Rham
cohomological operators in the derivation of the above nilpotent
symmetries in the framework of the superfield approach to BRST
formalism. One of our key future endeavors is to discuss and
derive the above nilpotent (anti-)co-BRST symmetry transformations
by imposing the appropriate restrictions on the six (4,
2)-dimensional supermanifold. Furthermore, the application of the
key ideas of our present investigation to the 4D gravitational
theories is another challenging project that we plan to pursue. In
this connection, we would like to point out that the {\it usual}
superfield formulation has already been applied to the 4D
gravitational theories to derive the (anti-)BRST symmetry
transformations for the gauge and (anti-)ghost fields [24]. In
this attempt, however, the (anti-)BRST symmetry transformations
associated with the matter fields of the gravitational theories
have not yet been obtained. We can apply the theoretical arsenals
of our newly proposed {\it augmented} superfield formulation [52,
53, 57-61] to derive the nilpotent symmetry transformations
associated with the matter fields. To derive the nilpotent
(anti-)co-BRST symmetry transformations for the gravitational
theory (in the framework of the superfield approach to BRST
formalism) is yet another direction for further investigation.
These are some of the promising issues that would be pursued in
the future and our results would be reported in our forthcoming
publications [62].\\

\begin{center}
{\bf Appendix A}
\end{center}

\noindent We demonstrate, in this Appendix, that the 2D
interacting 1-form Abelian U(1) gauge theory, where there is a
direct coupling between the U(1) gauge field $A_\mu$ and the
matter conserved current ($J_\mu = - e \bar\psi \gamma_\mu \psi$)
constructed with the help of Dirac fields, is a field theoretical
model for the Hodge theory. To this end in mind, let us begin with
the following 2D (anti-)BRST invariant Lagrangian density in the
Feynman gauge (see, e.g. [39,40]) $$
\begin{array}{lcl}
{\cal L}^{(m)}_B =  {\displaystyle {\cal B}\; E - \frac{1}{2}\;
{\cal B}^2 + B \;(\partial \cdot A) + \frac{1}{2}\; B^2 + \bar
\psi \bigl (i \gamma^\mu D_\mu - m \bigr ) \psi - i \partial_\mu
\bar C \partial^\mu C},
\end{array}\eqno(A.1)
$$ where $D_\mu \psi = \partial_\mu \psi + i e A_\mu \psi$ is the
covariant derivative on the Dirac field $\psi$ which results in
the interaction term ($- e \bar \psi \gamma^\mu A_\mu \psi$) in
the above Lagrangian density. The interaction term is primarily a
coupling between the U (1) gauge field $A_\mu$ and the conserved
matter current $J_\mu = - e \bar \psi \gamma^\mu \psi$ where
$\gamma^\mu$'s are the well known $2 \times 2$ Dirac matrices
\footnote{ We choose here the components of the Dirac matrices as
$\gamma^0 = \sigma_2, \gamma^1 = i \sigma_1$ so that the
$\gamma_5$ matrix is defined as: $ \gamma^5 = \gamma^0 \gamma^1
\equiv \gamma_5 = \sigma_3$. It can be checked that the following
relationships are satisfied:  $ \{\gamma^\mu, \gamma^\nu \} = 2
\eta^{\mu\nu}, \gamma_\mu \gamma_5 = \varepsilon_{\mu\nu}
\gamma^\nu$ where the 2D Levi-Civita antisymmetric tensor is
chosen to satisfy $\varepsilon_{01} = + 1 = \varepsilon^{10},
\varepsilon^{\mu\nu} \varepsilon_{\mu\lambda} = -
\delta^\nu_\lambda$, etc. In the above choices, the $\sigma$'s are
the usual Hermitian $2 \times 2$ Pauli matrices.}. The other
symbols carry the same meaning as the ones mentioned for the
Lagrangian density (2.2). The above Lagrangian density is endowed
with the following local, covariant, continuous, nilpotent (i.e.
$s_{(a)b}^2 = 0$) and anticommuting ($s_b s_{ab} + s_{ab} s_b =
0$) (anti-)BRST symmetry transformations $s_{(a)b}$ (see, e.g.
[39,40] for details) $$
\begin{array}{lcl}
&&s_b A_\mu = \partial_\mu C, \qquad s_b C = 0, \qquad s_b \bar C
= i B, \qquad s_b B = 0, \nonumber\\ && s_b E = 0, \qquad s_b
{\cal B} = 0, \qquad s_b (\partial \cdot A) = \Box C, \qquad s_b
F_{\mu\nu} = 0, \nonumber\\ && s_b \psi = - i\;e\; C\; \psi,
\qquad \;\;\; s_b \bar \psi = - i \; e\; \bar \psi\; C,
\nonumber\\ &&s_{ab} A_\mu = \partial_\mu \bar C, \qquad s_{ab}
\bar C = 0, \qquad s_{ab} C = - i B, \qquad s_{ab} B = 0,
\nonumber\\ && s_{ab} E = 0, \qquad s_{ab} {\cal B} = 0, \qquad
s_{ab} (\partial \cdot A) = \Box \bar C, \qquad s_{ab} F_{\mu\nu}
= 0, \nonumber\\ && s_{ab} \psi = - i \; e\; \bar C \; \psi,
\qquad s_{ab} \bar \psi = - i\; e\; \bar \psi\; \bar C,
\end{array} \eqno(A.2)
$$ because the above Lagrangian density transforms to  a total
derivative. In addition to the above nilpotent symmetry
transformations, there exists another set of local, covariant,
continuous, nilpotent (i.e. $s_{(a)d}^2 = 0$) and anticommuting
($s_d s_{ad} + s_{ad} s_d = 0$) (anti-)co-BRST symmetry
transformations $s_{(a)d}$ (see, e.g. [39,40] for details) $$
\begin{array}{lcl}
 &&s_d A_\mu = - \varepsilon_{\mu\nu}
\partial^\nu \bar C, \qquad s_d \bar C = 0, \qquad s_d  C = - i {\cal B}, \qquad
s_d {\cal B} = 0, \nonumber\\ && s_d E = \Box \bar C, \quad s_d B
= 0, \quad s_d (\partial \cdot A) = 0, \quad s_d F_{\mu\nu} =
[\varepsilon_{\mu\rho} \partial_\nu - \varepsilon_{\nu\rho}
\partial_\mu]\; \partial^\rho \bar C, \nonumber\\
&& s_d \psi = - i \; e\; \bar C\; \gamma_5\; \psi, \qquad s_d \bar
\psi = + \;i\; e \;\bar \psi \; \gamma_5 \; \bar C, \nonumber\\
&&s_{ad} A_\mu = - \varepsilon_{\mu\nu}
\partial^\nu  C, \qquad s_{ad}  C = 0, \qquad s_{ad} \bar C =
 i {\cal B}, \qquad s_{ad} {\cal B} = 0, \nonumber\\ && s_{ad} E
= \Box C, \quad s_{ad}  B = 0, \quad s_{ad} (\partial \cdot A) =
0, \quad s_{ad} F_{\mu\nu} = [\varepsilon_{\mu\rho} \partial_\nu -
\varepsilon_{\nu\rho}
\partial_\mu]\; \partial^\rho  C, \nonumber\\
&& s_{ad} \psi = - i \; e\;  C\; \gamma_5\; \psi, \qquad s_{ad}
\bar \psi = + \;i \;e \;\bar \psi \; \gamma_5 \;  C,
\end{array}\eqno(A.3)
$$ under which the above Lagrangian density (A.1) remains
quasi-invariant because it transforms to a total derivative. In
addition, the key requirement for the existence of the above
symmetry transformations is that the mass of the Dirac fields
should be zero (i.e.  $m = 0$). It should be noted that for the 4D
interacting U(1) gauge theory with Dirac fields, the
(anti-)co-BRST symmetry transformations are found to be non-local,
non-covariant, nilpotent and anticommuting (see. e.g. [30-35] for
details). It can {\it also} be noted that, for all the fields of
the Lagrangian density (A.1), the operator relationships $ [s_b
s_{ad} + s_{ad} s_b]\;\Omega = 0$ and $[s_d s_{ab} + s_{ab} s_d]\;
\Omega = 0$ are valid as can be checked by exploiting the
nilpotent (anti-)BRST and (anti-)co-BRST symmetry transformations
(A.2) and (A.3). Here $\Omega$ is the generic field of the
Lagrangian density (A.1) of the 2D interacting 1-form gauge
theory.

It is interesting to point out that the anticommutation relation
between the (anti-)BRST and (anti-)co-BRST symmetry
transformations (i.e. $s_w = s_b s_d + s_d s_b \equiv s_{ab}
s_{ad} + s_{ad} s_{ab}$) defines a bosonic symmetry $s_w$ (with
$s_w^2 \neq 0$). Under this symmetry transformation, the fields of
the Lagrangian density (A.1) transform as follows [39,40] $$
\begin{array}{lcl}
 &&s_w A_\mu = \partial_\mu {\cal B} + \varepsilon_{\mu\nu}
\partial^\nu B,  \qquad s_w E = - \Box B, \qquad s_w (\partial \cdot
A) = \Box {\cal B}, \nonumber\\ &&s_w  C = 0, \qquad s_w \bar C =
0, \qquad s_w {\cal B} = 0, \qquad s_{ad} B = 0, \nonumber\\ &&
s_w \psi = i\; e\; (\gamma_5 \; B - {\cal B}), \qquad s_w \bar
\psi = - i\; e\; (\gamma_5 \; B - {\cal B}).
\end{array}\eqno(A.4)
$$ The above transformations are also the symmetry transformations
for the Lagrangian density (A.1) (only in the case when the mass
of the Dirac particle is zero (i.e. $m = 0$)). Thus, the operator
algebra that is obeyed by all the transformation operators $s_r$
(with $r = b, ab, d, ad, w$) is exactly same as the one given in
(2.6). This demonstrates that all the de Rham cohomological
operators (cf. equation (2.7)) have found their analogue in terms
of the above transformation operators (and their generators) for
the 2D interacting 1-form Abelian U(1) gauge theory with Dirac
fields.

To establish that the above interacting U(1) gauge theory is a
tractable field theoretical model of the Hodge theory, we focus
now on the existence of the discrete symmetry transformations of
the theory and show their relevance to the Hodge duality $*$
operation of the differential geometry. It can be checked that the
following discrete transformations  $$
\begin{array}{lcl}
&& C \to \pm\; i\;\gamma_5\; \bar C, \qquad \bar C \to \pm\; i \;
\gamma_5\; C, \qquad B \to \mp\; i\; \gamma_5\; {\cal B}, \qquad
\psi \to \psi, \nonumber\\ && A_0 \to \pm\;i\;\gamma_5\; A_1,
\qquad A_1 \to \pm\;i\;\gamma_5 \; A_0, \qquad \bar \psi \to \bar
\psi, \qquad e \to \mp\; i\; e, \nonumber\\ && (\partial \cdot A)
\to \pm\; i\;\gamma_5\; E, \qquad E \to \pm \; i\; \gamma_5\;
(\partial \cdot A),
\end{array}\eqno(A.5)
$$ are the symmetry transformations for the Lagrangian density
(A.1) because the Lagrangian density remains invariant under it.
It should be noted that the above transformations are in the
matrix notation. That is to say, the transformations $C\to \pm i
\gamma_5 \bar C$, etc, imply that $C \hat {\bf 1} \to \pm i
\sigma_3 \bar C \hat {\bf 1}$ where $\hat {\bf 1}$ is the $2
\times 2$ unit matrix for the above 2D interacting 1-form U(1)
Abelian gauge theory. In other words, $C \to \pm i \gamma_5 \bar
C$ implies that the Lagrangian density remains invariant under the
transformations: $C \to \pm i \bar C$ and/or $C \to \mp i \bar C$.
In exactly similar manner, rest of the other transformations
should be interpreted. In fact, the sign flip in the above
transformations is taken care of by the rest of the analogous
transformations for the other fields of the theory so that {\it
together} they become the symmetry transformations of the
Lagrangian density (A.1) for the 1-form interacting Abelian gauge
theory.

As discussed earlier, the above discrete transformations are found
to be the analogue of the Hodge duality $*$ operation of the
differential geometry. To see it clearly, it can be seen that any
arbitrary generic field  $\Omega$ of the Lagrangian density (A.1)
has the transformation property $ * ( * \Omega) = \pm \Omega$
under the above two successive operations of the discrete symmetry
transformation where the $(+)$ sign stands for the generic field
$\Omega$ being $\psi$ and $\bar\psi$ and $(-)$ sign is meant for
all the rest of the bosonic fields of the theory. It is gratifying
to note that the analogue of the equation (2.13) now becomes $$
\begin{array}{lcl}
s_{(a)d}\; (\Omega) = \pm\; * \; s_{(a)b} \; * \; (\Omega).
\end{array}\eqno(A.6)
$$ The above equation shows that the interplay of the continuous
nilpotent symmetry transformations (cf. (A.2) and (A.3)) and the
discrete symmetry transformations (cf. (A.5)) look exactly
identical to the relationship $\delta = \pm * d *$ that exists
between the nilpotent co-exterior derivative $\delta (\delta^2 =
0)$ and the exterior derivative $d (d^2 = 0)$. In (A.6) too, the
$(+)$ sign stands for $\psi$ as well as $\bar\psi$ and the rest of
the fields of the theory correspond to $(-)$ sign. Moreover, the
presence of $*$ in the equation (A.6) corresponds to the discrete
symmetry transformations quoted in the equation (A.5).

It is worthwhile to point out that the $(\pm)$ signs in the
relationship $\delta = \pm * d *$ of the differential geometry
depend on (i) the dimensionality of the manifold on which the
Hodge duality $*$ operation is defined (for instance, for the even
dimensional manifold, it is always the minus sign that becomes
relevant [12,13]), and (ii) the inner product of the degree of the
forms that are involved in the definition of the duality. In fact,
it is due to the latter condition that, for the odd dimensional
manifold, the sign in the relation $\delta = \pm * d *$ can be
positive or negative (see. e.g., [12-16] for details). However, in
the relationship (A.6), the positive and negative signs depend on
the two successive operations of the discrete transformations
(A.5) on a particular field of the Lagrangian density (A.1). In
terms of the conserved charges $Q_r$ (with $r = b, ab, d , ad,
w$), corresponding to the symmetry transformations $s_r$ (with $r
= b, ab, d, ad, w)$, the Hodge decomposition theorem can be
defined in the quantum Hilbert space of states for the interacting
$U(1)$ gauge theory, too. Thus, we conclude that the 2D {\it
interacting} 1-form Abelian gauge theory is a tractable field
theoretical model for the Hodge theory because all the de Rham
cohomological operators of the differential geometry are defined
in terms of the {\it well-defined} symmetry transformations (and
their corresponding generators) for the Lagrangian density of the
above theory.

In the framework of the augmented superfield formulation, there
are ways to derive the nilpotent (anti-)BRST and (anti-)co-BRST
symmetry transformations for {\it all} the fields of the
Lagrangian density (A.1) of the theory [52-61]. The long-standing
problem of the derivation of the nilpotent (anti-)BRST and
(anti-)co-BRST symmetry transformations for the matter fields
$\psi$ and $\bar\psi$ of the interacting gauge theory, in the
framework of the superfield approach to BRST formalism, has been
resolved by taking recourse to (i) the equality of some conserved
quantities [52,53,57-61] on the appropriate dimensional
supermanifold (e.g. the four (2, 2)-dimensional supermanifold for
2D the gauge theory) on top of the horizontality condition, and
(ii) a single gauge invariant restriction on the matter
superfields of the supermanifold that owes its origin to the
(super) covariant derivatives and their connection with the
(super) curvature 2-forms [54-56]. However, we shall not elaborate
on these issues in our present endeavor because we have already
proven that the 2D interacting 1-form Abelian gauge theory is a
Hodge theory in the Lagrangian formulation.

\end{document}